\pdfoutput=1
\documentclass[11pt]{article}

\usepackage[final]{acl}

\usepackage{times}
\usepackage{latexsym}
 \usepackage{booktabs}
\usepackage{enumitem}
\usepackage[normalem]{ulem}

\usepackage[T1]{fontenc}
\usepackage[utf8]{inputenc}

\usepackage{microtype}

\usepackage{inconsolata}
\usepackage{amsmath}

\usepackage{graphicx}
\usepackage[ruled,vlined, ,linesnumbered]{algorithm2e}
\usepackage[table]{xcolor} 
\usepackage{tabularx}  
\usepackage{cleveref}
\title{\textsc{DMRetriever}: A Family of Models for Improved Text Retrieval in Disaster Management}

\author{
\textbf{Kai Yin}\textsuperscript{1} \quad 
\textbf{Xiangjue Dong}\textsuperscript{1} \quad 
\textbf{Chengkai Liu}\textsuperscript{1} \quad 
\textbf{Allen Lin}\textsuperscript{1} \\
\textbf{Lingfeng Shi}\textsuperscript{1} \quad 
\textbf{Ali Mostafavi}\textsuperscript{1} \quad 
\textbf{James Caverlee}\textsuperscript{1} \\
\textsuperscript{1}Texas A\&M University \\
\texttt{\{kai\_yin,xj.dong,liuchengkai,al001,lingfengs111,mostafavi,caverlee\}@tamu.edu} \\}

\begin{document}
\maketitle
\begin{abstract}

Effective and efficient access to relevant information is essential for disaster management. 
However, no retrieval model is specialized for disaster management, and existing general-domain models fail to handle the varied search intents inherent to disaster management scenarios, resulting in inconsistent and unreliable performance.
To this end, we introduce \textsc{DMRetriever}, the first series of dense retrieval models (33M to 7.6B) tailored for this domain. It is trained through a novel three-stage framework of bidirectional attention adaptation, unsupervised contrastive pre-training, and difficulty-aware progressive instruction fine-tuning, using high-quality data generated through an advanced data refinement pipeline. Comprehensive experiments 
demonstrate that \textsc{DMRetriever} achieves state-of-the-art (SOTA) performance across all six search intents at every model scale. Moreover, \textsc{DMRetriever} is highly parameter-efficient, with 596M model outperforming baselines over 13.3$\times$ larger and 33M model exceeding baselines with only 7.6\% of their parameters. All codes, data, and checkpoints are available at \href{https://github.com/KaiYin97/DMRETRIEVER}{this repository}.

\end{abstract}

\section{Introduction}

Disasters, both natural and man-made, inevitably disrupt communities, endanger lives, and cause severe economic losses \citep{fan2021disaster, liu2025artificial}. During such events, emergency responders, policymakers, and affected populations rely on information retrieval (IR) systems for decision-making, situational awareness, and coordination of relief efforts \citep{abbas2025exploring}. Timely and reliable information access is thus essential for effective disaster management \citep{priya2020taqe}. 

\begin{figure}[t]  
    \centering
    \includegraphics[width=\columnwidth]{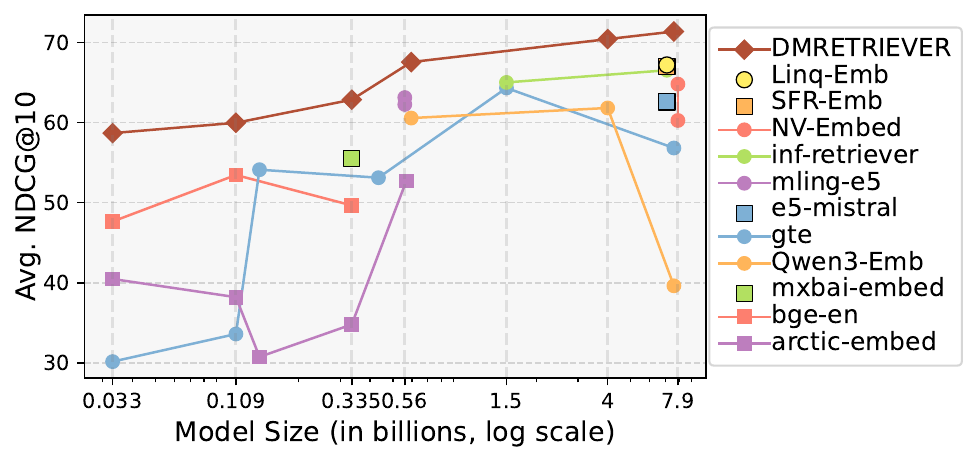}
    \caption{Average performance of \textsc{DMRetriever}. Breakdown for each search intent is in Table \ref{tab:embedding_results}}
    \label{fig:figure_1}
\vspace{-8pt}
\end{figure}

Information needs during disaster events are highly diverse, spanning fact-checking critical information, monitoring social media, and consulting technical or policy documents (See \S\ref{section: preliminary} for full lists of search intents) \citep{yin2025disastir}. However, a recent analysis of 30 popular general-domain IR models finds that \textit{none consistently achieve SOTA performance across all disaster management-specific search intents} \citep{yin2025disastir}. 
This limits their real-world applicability: some queries yield valuable insights, while others return irrelevant results, delaying decision-making and risking missed evidence that could compromise response efforts.

Furthermore, real-world disaster management demands \textit{scalable retrieval solutions}, from lightweight models for resource-constrained settings (e.g., on edge devices for rapid response) to larger models for performance-critical decision-making (e.g., in central command system). Hence, there is a great opportunity to \textbf{develop domain-specialized IR models of various scales that achieve both high performance and computational efficiency for disaster management.}

\begin{figure*}[t] 
    \centering
    \includegraphics[width=\textwidth]{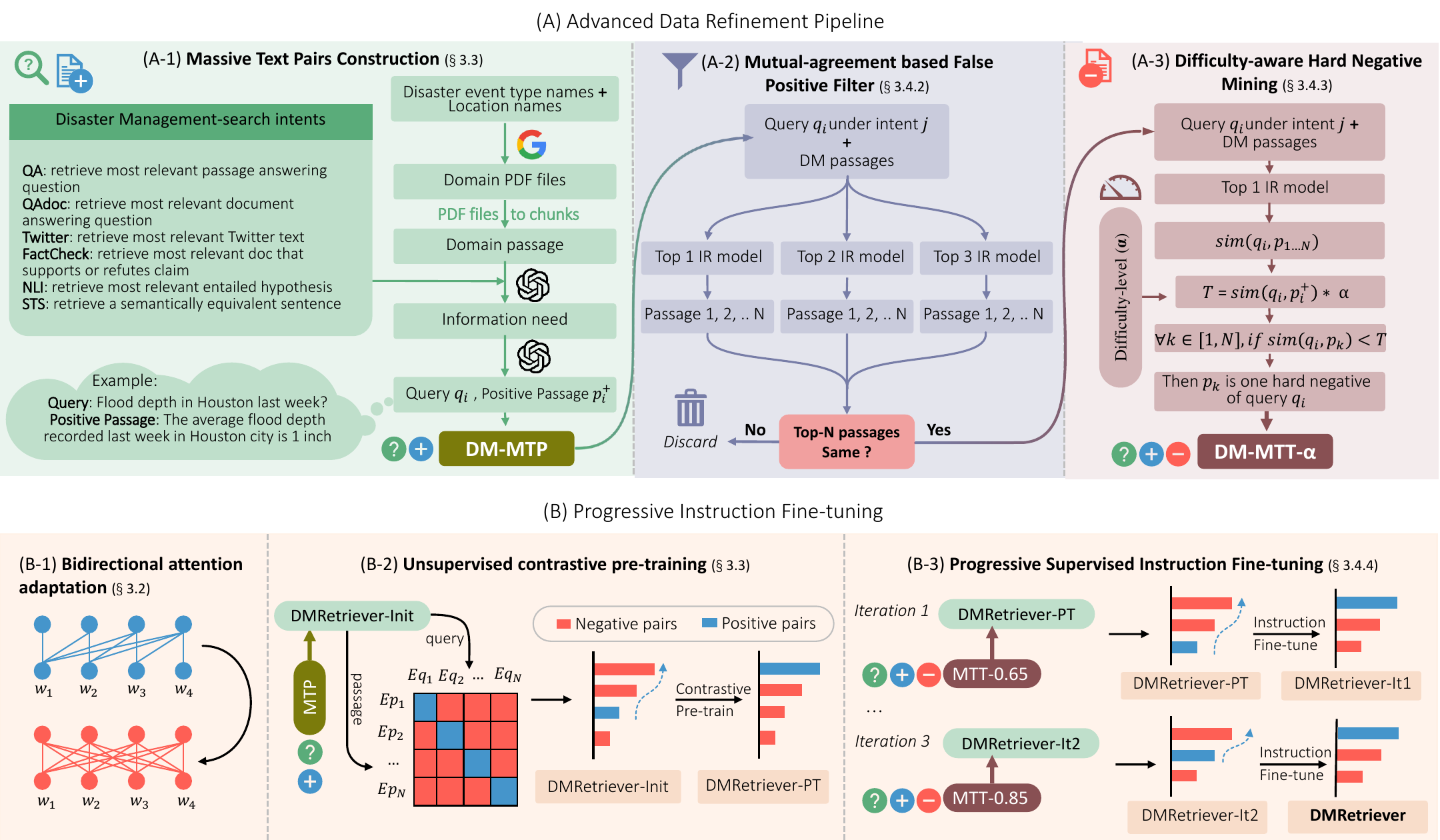}
    \caption{Overview of data refinement pipeline and progressive fine-tuning for \textsc{DMRetriever}. 
    MTP: \textbf{M}assive \textbf{T}ext \textbf{P}airs, MTT: \textbf{M}assive \textbf{T}ext \textbf{T}riplets.
    Both MTP and MTT in sub-figure (B) include DM (\textbf{D}isaster \textbf{M}anagement) and GD (\textbf{G}eneral \textbf{D}omain) variants, with constructions of GD variants omitted here and detailed in \S\ref{section: pt_dataset_construction} and \S\ref{sec: hard_neg_mine}. 
    DM passages include those extracted from PDFs and those generated by LLMs. DMRetriever-Init: backbone after bidirectional attention adaptation for decoder-only; original backbone for encoder-only. DMRetriever-PT: model after pre-training. DMRetriever-It2: model after iteration 2. 
    Algorithm~\ref{alg:dmretriever_train} details the full workflow.}
    \label{fig:workflow}
\vspace{-8pt}
\end{figure*}

To this end, we introduce \textsc{DMRetriever}, a family of dense retrieval models at six scales (33M–7.6B parameters) to improve retrieval performance in disaster management (Figure \ref{fig:figure_1}).

The success of domain-specific IR models in fields such as biomedicine \citep{xu2024bmretriever}, law \citep{chalkidis2020legal}, and finance \citep{sarmah2024hybridrag} shows the effectiveness of adding deep domain knowledge to improve performance. 
However, injecting disaster management knowledge into \textsc{DMRetriever} to improve performance poses \textbf{two challenges}: (1) the lack of high-quality, knowledge-rich data that covers diverse search intents during disasters, and (2) the absence of an efficient training method that can transfer such knowledge across models of different sizes.

To tackle the first challenge, we propose a \textbf{data refinement pipeline} (A in Figure \ref{fig:workflow}) to generate \emph{the first}, knowledge-rich data for disaster management. We first prompt a large language model (LLM)\footnote{We use GPT-4o-mini as LLM in this work.} with domain-specific passages to generate large-scale unlabeled query–passage pairs\footnote{\label{fn:definition}Following \citet{xiao2024c, wang2022text}, we treat data without mined negatives as unlabeled and data with mined hard negatives as labeled.} (A-1 in Figure \ref{fig:workflow}, \S\ref{section: pt_dataset_construction}). 
Given limitations of existing IR models in disaster management, we refine these noisy corpora with mutual-agreement filtering, which removes mislabeled pairs through model consensus (A-2 in Figure~\ref{fig:workflow}, \S\ref{sec: false_pos_filter}). Recognizing the varied learning capacities of different-sized \textsc{DMRetriever} models, we further propose difficulty-aware hard negative mining to generate negatives at varying difficulty levels (A-3 in Figure~\ref{fig:workflow}, \S\ref{sec: hard_neg_mine}), finally yielding a high-quality labeled dataset.

For the second challenge, we propose a \textbf{Progressive instruction fine-tuning} (B in Figure~\ref{fig:workflow}) that enables \textsc{DMRetriever} models of different sizes to effectively learn domain knowledge.
For large variants initialized from decoder-only backbones (Table~\ref{tab:DMretriever_models}), we adapt their causal attention into bidirectional attention, allowing access to information from future tokens (B-1 in Figure~\ref{fig:workflow}, \S\ref{section: train_S1}). Next, to better initialize \textsc{DMRetriever} and inject domain knowledge, we perform large-scale contrastive pre-training on unlabeled query–passage pairs (B-2 in Figure~\ref{fig:workflow}, \S\ref{section: train_S2}). Finally, to adapt \textsc{DMRetriever} to diverse search intents, we perform instruction fine-tuning on labeled data\footref{fn:definition}, where for small variants that have limited learning capacity, the difficulty of negative samples is progressively increased during fine-tuning (B-3 in Figure \ref{fig:workflow}, \S\ref{sec: prog_ft}).

Extensive experiments show that \textsc{DMRetriever} achieves new SOTA results on all six search intents at each scale (Figure \ref{fig:figure_1}). It is also highly parameter-efficient: 596M variant beats all XL-scale baselines while being over 13$\times$ smaller, and 33M variant surpasses all medium baselines with only 7.6\% of their parameters. Ablation study further confirms that knowledge generated by data refinement pipeline and injected through three-stage training framework substantially enhances model performance in disaster management.

The contributions of this work are as follows:

\begin{itemize}[topsep=0pt, partopsep=0pt, itemsep=0pt, parsep=0pt]
\item[(1)] We introduce \textsc{DMRetriever}, a family of six dense retrieval models (33M to 7.6B) that achieve new SOTA performance on each search intent across all scales, with superior parameter efficiency over existing baselines.
% \vspace{-5pt}

\begin{table*}[t]
\centering
\small
\setlength{\tabcolsep}{2pt}
\resizebox{\textwidth}{!}{%
\begin{tabular*}{\textwidth}{@{\extracolsep{\fill}}l|ccccc}
\hline
\textbf{Model} & \textbf{\#Parameters} & \textbf{Backbone} & \textbf{Backbone Type} & \textbf{Hidden Size} & \textbf{\#Layers} \\
\hline
\textsc{DMRetriever-33M}   & 33M   & MiniLM                         & encoder-only  & 384   & 12 \\
\textsc{DMRetriever-109M}  & 109M  & BERT-base-uncased             & encoder-only  & 768   & 12 \\
\textsc{DMRetriever-335M}  & 335M  & BERT-large-uncased-WWM        & encoder-only  & 1024  & 24 \\
\textsc{DMRetriever-596M}  & 596M  & Qwen3-0.6B                    & decoder-only  & 1024  & 28 \\
\textsc{DMRetriever-4B}    & 4B    & Qwen3-4B                      & decoder-only  & 2560  & 36 \\
\textsc{DMRetriever-7.6B}   & 7.6B  & Qwen3-8B                      & decoder-only  & 4096  & 36 \\
\hline
\end{tabular*}}
\caption{Model configurations of \textsc{DMRetriever} family
% , initialized from encoder-only and decoder-only backbones 
across six model sizes. 
\textsc{DMRetriever}-7.6B is initialized from Qwen3-8B after removing its language modeling head, leaving 7.568B parameters (rounded to 7.6B).
}
\label{tab:DMretriever_models}
\vspace{-5pt}
\end{table*}

\item[(2)] We propose a data refinement pipeline 
% with massive text pairs construction, mutual-agreement-based filtering, and difficulty-aware hard negative mining, 
yielding the first large-scale training dataset for disaster management. It is generalizable to other domains with limited annotations.
% \vspace{-5pt}

\item[(3)] We introduce progressive instruction fine-tuning 
% that combines bidirectional adaptation, large-scale unsupervised contrastive pre-training, and progressive supervised fine-tuning, 
which enables models of varied sizes to effectively absorb domain knowledge and enhances their ability across different search intents.
\end{itemize}

\section{Related Work}

General-domain retrieval has been long dominated by bidirectional encoders such as BERT \citep{devlin2019bert} and T5 \citep{raffel2020exploring}, with models like Sentence-BERT \citep{reimers2019sentence} and SimCSE \citep{gao2021simcse} improving performance through contrastive pre-training and supervised fine-tuning on MS MARCO \citep{nguyen2016ms}. More recently, decoder-only models, from GPT-based embeddings \citep{brown2020language} to models such as Qwen3-Embedding \citep{qwen3embedding}, have leveraged instruction tuning and large-scale synthetic data to achieve SOTA results.

However, these general-domain models face clear limitations in disaster management, with no approach achieving SOTA performance across diverse search intents \citep{yin2025disastir} due to distribution shift issue \citep{thakur2021beir}. Prior work on disaster information access has mainly targeted on social media, emphasizing clustering of posts, event or situational classification, and named entity recognition \citep{imran2015processing, alam2021crisisbench, yin2024crisissense}. While valuable, these efforts remain narrow in scope and have yet to address the need for robust retrieval systems tailored to disaster management-related diver search intents.

\section{\textsc{DMRetriever}}

\textsc{DMRetriever} is a family of dense retrieval models at six scales (33M–7.6B) 
initialized from encoder- and decoder-only backbones (Table~\ref{tab:DMretriever_models}). This scalability supports real-world disaster management applications ranging from resource-constrained settings with smaller models to performance-critical scenarios with larger ones.

% \textsc{DMRetriever} is a family of dense retrieval models spanning six parameter scales: 33M, 109M, 335M, 596M, 4B, and 7.6B, initialized from both encoder-only and decoder-only Transformer backbones (Table \ref{tab:DMretriever_models} for model details). The scalability across model sizes enables DMRetriever to support a wide range of disaster management information retrieval applications, from resource-constrained deployments with smaller models to performance-critical scenarios requiring larger ones.

\subsection{Preliminaries}
\label{section: preliminary}

In \textsc{DMRetriever}, queries and passages are encoded into dense vectors using a shared embedding model $\mathbf{E}(\cdot)$. Given a query $q$ and a passage $p$, we prepend a search intent-specific instruction $I_q$ to the query to improve search intent awareness \citep{wei2021finetuned, ouyang2022training, lee2024nv}. 

Following DisastIR \citep{yin2025disastir}, we consider six disaster management-specific search intents: question-answer (QA), Twitter (TW), Fact Checking (FC), Natural Language Inference (NLI), and Semantic Textual Similarity (STS). Table \ref{tab:PT_instruction_formats} details corresponding instruction formats.

% Following \citet{yin2025disastir}, we choose six different types of search tasks in disaster management areas: question-answer (QA) retrieval, Twitter retrieval, Fact Checking (FC) retrieval, Natural Language Inference (NLI) retrieval, and Semantic Textual Similarity (STS) Retrieval. Appendix \ref{appendix:search_task} details these search tasks along with the corresponding instruction format. 

The embeddings are then computed as $\mathbf{e}_q = \mathbf{E}(I_q \oplus q)$ and $\mathbf{e}_p = \mathbf{E}(p)$, where $\oplus$ denotes sequence concatenation. We apply mean pooling over the final hidden states to obtain fixed-length representations. The relevance score $\text{sim}(q, p)$ is computed as the cosine similarity between embeddings, scaled by a temperature parameter $\tau$:
\begin{equation}
\text{sim}(q, p) = \cos(\mathbf{e}_q, \mathbf{e}_p) / \tau.
\label{eq:similarity}
\end{equation}

To effectively adapt \textsc{DMRetriever} to disaster management domain, we propose a three-stage training framework using high-quality training data produced from our advanced data refinement pipeline (Figure \ref{fig:workflow} and Algorithm \ref{alg:dmretriever_train}).

% : (1) enabling bidirectional attention for decoder-only backbones(\S\ref{section: train_S1}), (2) unsupervised contrastive pre-training with in-batch negatives(\S\ref{section: train_S2}), and (3) difficulty-aware progressive supervised instruction fine-tuning(\S\ref{section: train_S3}).

\subsection{Enabling Bidirectional Attention in Decoder-only Backbones}
\label{section: train_S1}

\subsubsection{Bidirectional attention adaptation} 
For \textsc{DMRetriever} initialized from decoder-only backbones, we replace the causal attention mask with an all-ones matrix (B-1 in Figure~\ref{fig:workflow}) to enable the model to capture information from future tokens, inspired by the recent success of \citet{lee2024nv, behnamghader2024llm2vec, muennighoff2024generative}. This replacement allows each token to attend to every other token in the sequence, converting it into a bidirectional model. 

To make model aware of its bidirectional attention, we train model by using masked language model (MLM) training objective \citep{devlin2019bert}. Given an input sequence, we randomly mask some of tokens and then train model to predict masked tokens based on past and future tokens.

\subsubsection{MLM training data construction} 
We perform MLM training using a mix of general-domain and disaster management corpora. For the general domain, we use the English Wikipedia from \citet{merity2016pointer}. For the disaster management domain, we use passages from DM-MTP (constructed in \S\ref{section: pt_dataset_construction}).

% following \citet{yin2025disastir}, we crawl the web with 301 disaster event types as queries, collect domain-specific PDFs, and process them into clean, high-quality passages. Appendix~\ref{appendix:MLM_set} provides details of MLM training data construction. 

% [Remove passages contained in the test set by exact string match, although the corpus is not considered as contamination. xxx (need do add this)]

\subsection{Unsupervised contrastive pre-training}
\label{section: train_S2}

\subsubsection{Pre-training dataset construction}
\label{section: pt_dataset_construction}

% We construct MTP-PT (\textbf{M}assive \textbf{T}ext \textbf{P}airs--\textbf{P}re-\textbf{T}raining; See Algorithm~\ref{alg:dmrpt} for its construction.) for pre-training. It comprises two large-scale subsets: GD-MTP-PT (\textbf{G}eneral-\textbf{D}omain--\textbf{M}assive \textbf{T}ext \textbf{P}airs--\textbf{P}re-\textbf{T}raining) for the general domain and DM-MTP-PT (\textbf{D}isaster \textbf{M}anagement--\textbf{M}assive \textbf{T}ext \textbf{P}airs--\textbf{P}re-\textbf{T}raining) for the disaster management domain.

% To equip \textsc{DMRetriever} with a basic understanding of disaster management contexts, 
To robustly initialize \textsc{DMRetriever} for retrieval tasks and inject domain knowledge, we construct MTP (\textbf{M}assive \textbf{T}ext \textbf{P}airs; see Algorithm~\ref{alg:dmrpt}), which consists of two large-scale subsets: DM-MTP (disaster management domain) and GD-MTP (general-domain). Its composition is shown in Table~\ref{tab:PT_data}.

% For the disaster management domain, no publicly available paired datasets exist. To address this gap, we construct DM-MTP, \textbf{the first} large-scale synthetic training dataset of query--passage pairs tailored for disaster management retrieval tasks. 

In the absence of paired datasets for disaster management, DM-MTP represents the first large-scale collection of query–passage pairs in this domain. Inspired by \citet{wang2023improving} and \citet{yin2025disastir}, we propose a three-stage pipeline to generate it (A-1 in Figure \ref{fig:workflow}, lines 1--7 in Algorithm \ref{alg:dmrpt}).
% few-shot prompting strategy~\citep{yin2025disastir} covering six predefined search intents. 
% In the first stage,  we crawl the web using 301 disaster event types as queries, collect domain-specific PDFs, and process them into clean, high-quality passages. Secondly, for each disaster management passage, an LLM generates information-need statements grounded in the passage’s content. A randomly selected information-need statement is paired with the passage to prompt LLM to produce a user query and a directly relevant passage. The relevant passage is treated as positive passage for the generated user query in DM-MTP. 
In the first stage, we crawl the web with queries formed by combining disaster event type names and location names, collect domain-specific PDFs, and process them into chunks, taking these as disaster-domain passages. Secondly, for each passage, a LLM generates information-need statement taking passage content and its designated search intent definition as input. Finally, the information-need statement is paired with the passage to prompt the LLM to generate a user query and a directly relevant passage, which serves as the positive passage for this query in DM-MTP. Appendix~\ref{appendix:DM-MTP} details the whole DM-MTP construction process.

To enrich \textsc{DMRetriever} with diverse linguistic and semantic relevance patterns, GD-MTP is built by aggregating 15 publicly available datasets (Table~\ref{tab:PT_data}). Query--positive passage pairs are created from unlabeled data using search intent-specific heuristics (Lines 8--12 in Algorithm \ref{alg:dmrpt}). For example, in datasets associated with QA intents, user questions are treated as queries and their answers as positive passages. Heuristics for other search intents are described in Appendix~\ref{appendix:PT_set}.

For negative sampling in MTP, instead of explicitly mining hard negatives, we adopt the in-batch negative strategy~\citep{karpukhin2020dense}, where passages from other pairs within the same mini-batch serve as negative examples.

\subsubsection{Contrastive Pre-training} 
\label{sec: PT}
\textsc{DMRetriever} is pre-trained on MTP using contrastive learning to separate relevant query–passage pairs from irrelevant ones. For each mini-batch $\mathcal{B}$, we adopt InfoNCE loss \citep{chen2020simple} as pre-training objective, encouraging similarity between $q_i$ and its positive passage $p_i^+$ to be higher than that with any other passage $p_j \in \mathcal{B},\, j \neq i$ within the same batch (B-2 in Figure~\ref{fig:workflow}):

\begin{equation}
\mathcal{L}_{\text{cpt}} = 
-\log 
\frac{e^{\mathrm{sim}(q_i, p_i^+)}}
{\sum\limits_{p_j \in \mathcal{B}} e^{\mathrm{sim}(q_i, p_j)}}.
\label{eq:cpt_infonce}
\end{equation}

% Contrastive pre-training improves embedding quality by aligning semantically similar texts and separating dissimilar ones, enabling the retriever to better adapt to retrieval tasks in the disaster management domain.

% \subsection{Difficulty-aware Supervised Contrastive Fine-tuning}
\subsection{Difficulty-aware Progressive Supervised Instruction Fine-tuning}
\label{section: train_S3}

\subsubsection{Fine-tuning dataset overview}
\label{section: FT_dataset}
To further enhance domain-specific retrieval capabilities, \textsc{DMRetriever} is fine-tuned on a labeled dataset, MTT (\textbf{M}assive \textbf{T}ext \textbf{T}riplets; Algorithm~\ref{alg:dmrft}) comprising: DM-MTT for disaster management domain and GD-MTT for general domain. DM-MTT is obtained by applying false positive filtering (\S\ref{sec: false_pos_filter}) and hard negative mining (\S\ref{sec: hard_neg_mine}) to query–passage pairs in DM-MTP, while pairs in GD-MTP undergo hard negative mining to produce GD-MTT. Its composition is listed in Table \ref{tab:FT_data}.

\subsubsection{Mutual-agreement False Positive Filter}
\label{sec: false_pos_filter}

Previous studies \citep{alberti2019synthetic, xu2024bmretriever, wang2022text} show that LLM synthetic data can be noisy, and they commonly adopt a consistency-based filtering (CBF for short) approach: discarding pairs in which LLM-generated positive passage is not ranked within the top-$k$ results returned by a reference embedding model. 

% In disaster management domain, since no existing open-source model achieves the best performance across diverse search intents \citep{yin2025disastir}, 
Given the limited capability of existing open-source IR models in disaster management, we propose a \emph{mutual-agreement-based false positive filtering} method to more effectively remove low-quality pairs from DM-MTP (A-2 in Figure~\ref{fig:workflow}, lines~1--8 in Algorithm~\ref{alg:dmrft}). For each query, we use the top three open-source IR models identified in DisastIR for its corresponding search intent (Table~\ref{tab:ir-top-models}) to retrieve the top-$N$ passages. A query is retained only if all three models return the same top-$N$ passages; in that case, its top-1 passage is designated as the positive passage of this query.

% Section \ref{sec: ablation_false_filter} compares our method with CBF method.

% In Section \ref{section: abalation}, we compare the effectiveness of our mutual-agreement filter with the widely used consistency-based filtering method.

\subsubsection{Difficulty-aware Hard Negative Mining}
\label{sec: hard_neg_mine}
% After mutual-agreement filtering, each remaining 
Query–positive passage pairs further undergo hard negative mining to provide challenging negatives for enhancing \textsc{DMRetriever}'s performance. Given that \textsc{DMRetriever} spans a wide range of model parameter sizes with different learning capacities, we propose a \emph{difficulty-aware hard negative mining strategy} that caters to each model’s learning ability by providing negatives at appropriate difficulty levels.

% We identify hard negatives by comparing each candidate passage's similarity score with that of positive passage \citep{moreira2024nv}.
Inspired by \citet{moreira2024nv}, for a query \(q_i\) with a positive passage \(p_i^+\), a passage \(p_k\) is treated as a hard negative \(p_{i,k}^-\) if  
\begin{equation}
\mathrm{sim}(q_i,p_k) < \mathrm{sim}(q_i,p_i^+) \times \alpha.
\label{eq:psnr_criterion}
\end{equation}
For each query $q_i$, we employ the best-performing IR model corresponding to its search intent (See Table~\ref{tab:ir-top-models}) to retrieve the top 200 passages from the associated passage corpus. From this pool, we select the top-$K$ passages that satisfy Equation~\ref{eq:psnr_criterion}, forming the hard negative set $\mathcal{H}_i$ (A-3 in Figure~\ref{fig:workflow},  Lines~10--15 in Algorithm~\ref{alg:dmrft}). 

% This margin $\alpha$ serves as a tunable parameter that controls the difficulty of mined negatives. Instead of using the same difficulty level for different variants of \textsc{DMRetriever}, we propose to adjust the difficulty level of mined hard negative for model variants in different sizes considering their different learning ability. This allows us to generate datasets tailored to specific difficulty levels, which we further denote as MTT-$\alpha$.

The margin $\alpha$ serves as a tunable parameter that controls the difficulty of mined negatives, referred to as the \emph{difficulty level}.
Instead of applying a uniform difficulty level across all \textsc{DMRetriever} variants, we adjust $\alpha$ based on model size to better align with their learning capacities.
This yields datasets tailored to specific difficulty levels, denoted as MTT-$\alpha$.

\begin{table*}[ht]
\centering
\small
\resizebox{\textwidth}{!}{
\begin{tabular*}{\textwidth}{@{\extracolsep{\fill}}l|c|cccccc|c@{\hspace{5pt}}}
\hline
\textbf{Model} & \textbf{Scale} & \textbf{QA} & \textbf{QAdoc} & \textbf{TW} & \textbf{FC} & \textbf{NLI} & \textbf{STS} & \textbf{Avg.} \\
\hline
\rowcolor{lightgray!50}
\multicolumn{9}{l}{\textbf{Small Size ($\leq$109M)}} \\
\hline
thenlper-gte-small \citep{li2023gte}  & 33M & 18.04 & 9.13 & 10.95 & 49.63 & 37.51 & 55.55 & 30.14    \\
arctic-embed-m \citep{merrick2024arctic}  & 109M & 33.15 & 14.04 & 8.48 & 35.07 & 38.67  & 56.20 & 30.94  \\
thenlper-gte-base \citep{li2023gte} & 109M & 9.18 & 5.42 & 37.91 & 60.45 & 42.52  & 46.07 & 33.59  \\

arctic-embed-m-v1.5 \citep{merrick2024arctic}     & 109M & 25.76 & 30.41 & 17.95 & 47.97 & 42.88 & 64.16 & 38.19 \\
arctic-embed-s \citep{merrick2024arctic}        & 33M  & 38.58 & 28.81 & 21.33 & 47.21 & 39.85  & 66.96 & 40.46 \\
bge-small-en-v1.5 \citep{xiao2024c}                & 33M  & 56.91 & 51.19 & 25.15 & 55.17 & 32.87  & 64.54 & 47.64 \\
bge-base-en-v1.5  \citep{xiao2024c}    & 109M & 51.50 & 52.78 & 46.72 & 59.93 & 41.16  & \underline{68.63} & 53.45 \\

\rowcolor{green!20}
\textsc{DMRetriever-33M} (ours) & 33M & \underline{62.47}\textsuperscript{†} & \underline{57.03}\textsuperscript{†} & \underline{57.22}\textsuperscript{†} & \underline{60.81}\textsuperscript{†} & \underline{46.56}\textsuperscript{†} & 67.57 & \underline{58.61}\textsuperscript{†} \\
\rowcolor{green!20}
\textsc{DMRetriever-109M}   (ours)               & 109M & \textbf{63.19}\textsuperscript{†} & \textbf{59.55}\textsuperscript{†} & \textbf{58.88}\textsuperscript{†} & \textbf{62.48}\textsuperscript{†} & \textbf{46.93}\textsuperscript{†} & \textbf{68.79}\textsuperscript{†} & \textbf{59.97}\textsuperscript{†} \\
\hline
\rowcolor{lightgray!50}
\multicolumn{9}{l}{\textbf{Medium Size (137M--335M)}} \\
\hline
arctic-embed-m-long   \citep{merrick2024arctic}   & 137M & 21.51 & 10.86 & 19.24 & 36.13 & 41.67 & 54.94 & 30.73   \\
arctic-embed-l \citep{merrick2024arctic} & 335M & 40.56 & 30.19 & 14.98 & 32.64 & 34.20  & 56.10 & 34.78  \\
bge-large-en-v1.5  \citep{xiao2024c}                & 335M & 56.76  & 54.45 & 32.20 & 54.90 & 35.11 &  64.47 & 49.65 \\

gte-base-en-v1.5   \citep{li2023gte}   & 137M & 60.51 & 55.62 & 46.26 & 52.24 & 39.59 &  \underline{70.40} & 54.10 \\
mxbai-embed-large-v1  \citep{emb2024mxbai}              & 335M & \underline{64.24} & \underline{62.63} & 39.94 & \underline{58.12} & 40.18  & 68.01 & 55.52 \\
arctic-embed-m-v2.0 \citep{merrick2024arctic}      & 305M & 61.22 & 62.20 & \underline{47.01} & 57.79 & \underline{42.29} & 64.51 & \underline{55.84} \\

\rowcolor{green!20}
\textsc{DMRetriever-335M} (ours)        & 335M & \textbf{67.44}\textsuperscript{†} & \textbf{62.69}\textsuperscript{†} & \textbf{62.16}\textsuperscript{†} & \textbf{64.42}\textsuperscript{†} & \textbf{49.69}\textsuperscript{†} & \textbf{70.71}\textsuperscript{†} & \textbf{62.85}\textsuperscript{†} \\
\hline
\rowcolor{lightgray!50}
\multicolumn{9}{l}{\textbf{Large Size (434M--1.5B)}} \\
\hline
arctic-embed-l-v2.0 \citep{merrick2024arctic}     & 568M & 55.23 & 59.11 & 38.11 & 60.10 & 41.07 & 62.61 & 52.70 \\

gte-large-en-v1.5  \citep{li2023gte}                 & 434M & 67.37 & 58.18 & 39.43 & 52.66 & 34.45 & 66.47 & 53.09 \\

Qwen3-Embedding-0.6B  \citep{qwen3embedding}             & 596M & 66.10  & 52.31  & 62.38  & 64.89 & 50.30 & 67.39 & 60.56 \\
mulling-e5-large-instruct \citep{wang2024multilingual}    & 560M & 67.97 & \underline{64.64} & 62.25 & \underline{66.78} & 48.51 & 63.42 & 62.26 \\
mulling-e5-large \citep{wang2024multilingual}             & 560M & 66.99 & 64.01 & 62.81 & 59.87 & 50.93 & \underline{74.12} & 63.12 \\
gte-Qwen2-1.5B-instruct \citep{li2023gte}  & 1.5B & \underline{69.85} & 59.17 & \underline{65.09} & 62.73 & \underline{55.51}  & 73.58 & 64.32 \\
inf-retriever-v1-1.5b  \citep{infly-ai_2025}            & 1.5B & 69.41 & 64.29 & 62.99 & 65.39 & 54.03 & 73.92 & \underline{65.01} \\

\rowcolor{green!20}
\textsc{DMRetriever-596M} (ours)         & 596M & \textbf{72.44}\textsuperscript{†} & \textbf{67.50}\textsuperscript{†} & \textbf{65.79}\textsuperscript{†} & \textbf{69.15}\textsuperscript{†} & \textbf{55.71}\textsuperscript{†} & \textbf{74.73}\textsuperscript{†} & \textbf{67.55}\textsuperscript{†} \\
\hline
\rowcolor{lightgray!50}
\multicolumn{9}{l}{\textbf{XL Size ($\geq$4B)}} \\
\hline
Qwen3-Embedding-8B   \citep{qwen3embedding}              & 7.6B & 44.21 & 34.38 & 41.56 & 42.04 & 32.53 & 42.95 & 39.61 \\
gte-Qwen2-7B-instruct \citep{li2023gte} & 7.6B & 70.24 & 47.41 & 63.08 & 31.62 & 53.71  & 74.88 & 56.82 \\
NV-Embed-v1   \citep{moreira2024nv}                     & 7.9B & 68.06 & 62.70 & 56.02 & 59.64 & 48.05 & 67.06 & 60.26 \\
Qwen3-Embedding-4B   \citep{qwen3embedding}              & 4B   & 67.20  & 59.14  & 65.28  & 67.16 & 53.61 & 58.51 & 61.82 \\
e5-mistral-7b-instruct \citep{wang2023improving} & 7.1B  & 65.57 & 64.97 & 63.31 & 67.86 & 47.55 & 66.48 & 62.58   \\

NV-Embed-v2   \citep{lee2024nv}                     & 7.9B & 74.47 & 69.37 & 42.40 & 68.32 & \underline{58.20} & 76.07 & 64.80 \\

inf-retriever-v1  \citep{infly-ai_2025}                  & 7.1B & 72.84 & 66.74 & 66.23 & 65.53 & 51.86 & 75.98 & 66.53 \\
SFR-Embedding-Mistral  \citep{SFRAIResearch2024}            & 7.1B & 71.41 & 67.14 & 69.45 & 70.31 & 50.93 & 72.67 & 66.99 \\
Linq-Embed-Mistral \citep{LinqAIResearch2024}  & 7.1B & 74.40 & 70.31 & 64.11 & 70.64 & 52.46  & 71.25 & 67.19 \\
\rowcolor{green!20}
\textsc{DMRetriever-4B} (ours)            & 4B   & \underline{75.32}\textsuperscript{†} & \underline{70.23}\textsuperscript{†} & \underline{70.55}\textsuperscript{†} & \underline{71.44}\textsuperscript{†} & 57.63 & \underline{77.38}\textsuperscript{†} & \underline{70.42}\textsuperscript{†} \\
\rowcolor{green!20}
\textsc{DMRetriever-7.6B} (ours)         & 7.6B & \textbf{76.19}\textsuperscript{†} & \textbf{71.27}\textsuperscript{†} & \textbf{71.11}\textsuperscript{†} & \textbf{72.47}\textsuperscript{†} & \textbf{58.81}\textsuperscript{†} & \textbf{78.36}\textsuperscript{†} & \textbf{71.37}\textsuperscript{†} \\
\hline
\end{tabular*}
}
\caption{Experiment results across all six search intents at various scales in DisastIR-Test. 
\textbf{Bold} and \underline{underline} denote the best and second-best scores within each size group. \textsuperscript{†} indicates
improvement over the best baseline is statistically significant with $p$-value $< 0.05$ evaluated using the one-tailed Wilcoxon signed-rank test.}

\label{tab:embedding_results}
\vspace{-5pt}
\end{table*}

\begin{table}[ht]
\centering
\small
\setlength{\tabcolsep}{3.3pt} 
\resizebox{\columnwidth}{!}{
\begin{tabular}{c|cccccc|c}
\hline
\textbf{Size} & \textbf{QA} & \textbf{QAdoc} & \textbf{TW} & \textbf{FC} & \textbf{NLI} & \textbf{STS} & \textbf{Avg} \\
\hline
33M   & 57.72 & 55.83 & 57.40 & 59.17 & 41.55 & 64.71 & 56.06 \\
109M  & 59.41 & 58.24 & 58.72 & 60.68 & 42.13 & 62.93 & 57.02 \\
335M  & 64.10 & 61.36 & 62.09 & 63.33 & 44.86 & 66.55 & 60.38 \\
596M  & 71.39 & 67.00 & 67.23 & 68.71 & 49.91 & 71.66 & 65.98 \\
4B    & \underline{72.35} & \underline{68.68} & \textbf{70.06} & \underline{71.09} & \underline{53.69} & \underline{74.47} & \underline{68.39} \\
7.6B  & \textbf{73.47} & \textbf{69.77} & \underline{69.42} & \textbf{71.44} & \textbf{53.91} & \textbf{75.56} & \textbf{68.93} \\
\hline
\end{tabular}
}
\caption{Performance of \textsc{DMRetriever-PT}.}
\label{tab:unsup_models}
\vspace{-7pt}
\end{table}

\subsubsection{Progressive Supervised Instruction Fine-tuning}
\label{sec: prog_ft}

Previous studies show that retrieval performance degrades when trained with negatives that are either too difficult or too simple \citep{merrick2024arctic}. To address this, for small-sized variants of \textsc{DMRetriever} (33M, 109M, and 335M) that have limited learning capacity after pre-training, we introduce \emph{progressive instruction fine-tuning} (B-3 in Figure~\ref{fig:workflow}, Lines~9--13 in Algorithm~\ref{alg:dmretriever_train}), inspired by curriculum learning \citep{bengio2009curriculum}. This approach gradually increases the difficulty of training samples across iterations,\footnote{An iteration refers to a curriculum stage in Lines 9–12 of Algorithm~\ref{alg:dmretriever_train}, each consisting of multiple training batches.} where the best validation checkpoint from one iteration initializes the next. At each stage, the difficulty level $\alpha$ is raised to make negatives harder, with models progressively trained on MTT-0.65, 0.75, and 0.85.

In contrast, the larger variants (596M, 4B, and 7.6B), which already exhibit strong performance after pre-training, are fine-tuned only on MTT-0.95. $\alpha$ value selections are detailed in Appendix~\ref{appendix:alpha_choice}.

To further adapt \textsc{DMRetriever}'s embeddings to different search intents, we perform instruction fine-tuning, where each query $q_i$ is prepended with an intent-specific instruction $I_q$, forming the input $\tilde{q}_i = [I_q; q_i]$. The model is then fine-tuned on MTT-$\alpha$ at each iteration using the InfoNCE loss: 
\begin{equation}
\mathcal{L}_{\text{ft}} = 
- \log 
\frac{e^{\mathrm{sim}(\tilde{q}_i, p_i^+)}}
{e^{\mathrm{sim}(\tilde{q}_i, p_i^+)} 
+ \sum\limits_{p_{i,k}^- \in \mathcal{H}_i} 
e^{\mathrm{sim}(\tilde{q}_i, p_{i,k}^-)}}.
\label{eq:hnm_infonce}
\end{equation}

\section{Experiment Setup}

% \subsection{Experiment Setup}
\subsection{DisastIR-DevLite and DisastIR-Test Construction}
% We partition DisastIR \footnote{DisastIR is the most comprehensive IR benchmark for disaster management available at \href{https://huggingface.co/datasets/KaiYinTAMU/DisastIR}{this repository}.} into two subsets: \textbf{DisastIR-DevLite}, used as a lightweight validation set, and \textbf{DisastIR-Test}, used as a test set. \textsc{DMRetriever} is evaluated on DisastIR-Test. Table \ref{tab:disastir-stats} details statistics of these two subsets. \textsc{DMRetriever} training is detailed in Appendix \ref{appendix:DM_train}.

% To develop validation set for model development, prior works \citep{boteva2016full, yang2018hotpotqa, thakur2021beir} randomly split queries into validation and test sets, while \emph{keeping passage corpus identical for both}. However, encoding the entire collection of passages in the corpus during model development can be computationally expensive. 

Prior works \citep{boteva2016full, yang2018hotpotqa, thakur2021beir} typically construct validation sets by randomly splitting queries while keeping the passage corpus identical across validation and test sets, which makes repeatedly encoding the full passage corpus computationally expensive. 

To address this, we introduce \textbf{DisastIR-DevLite} (See Algorithm \ref{alg:disastir_devlite}), a lightweight validation set derived from DisastIR\footnote{DisastIR is the only publicly released IR benchmark for disaster management available at \href{https://huggingface.co/datasets/DMIR01/DisastIR}{this repository}.} by sampling queries and constructing a much smaller passage corpus tailored to them, while the remaining queries form \textbf{DisastIR-Test} with the full DisastIR passage corpus. Full construction details are provided in Appendix~\ref{appendix:disastir_devlite}.

DisastIR-DevLite contains only 3.9 \% of the original passages (Table~\ref{tab:disastir-stats}), yet remains challenging and informative for model development. Appendix~\ref{appendix:disastir_dev} highlights its practicality, showing that DevLite enables over 30$\times$ faster model development while maintaining reliable performance rankings (Kendall’s $\tau > 0.90$), thus supporting rapid yet trustworthy experimentation.

\subsection{Baselines and Metrics} 
\label{sec: baselines}
Baselines are drawn from two sources:  (1) models evaluated in DisastIR, and (2) the latest high-performing open-source models on the MTEB retrieval benchmark \footnote{As of August 15, 2025.} \citep{muennighoff2022mteb} that have not been evaluated on DisastIR. 

Some baselines in DisastIR are fine-tuned with knowledge distillation (KD), which uses external supervision from a teacher model. In such cases, models are not trained solely on ground-truth data but are also guided by teacher’s high-quality interpretation of it. We thus exclude these KD-based models as unfair comparisons, leaving 29 baselines ranging from 33M to 7.9B parameters. Detailed reasons for this exclusion are in Appendix~\ref{appendix:remove_kd}. Baseline implementations are detailed in Appendix~\ref{appendix:eva_models}.

Model performance is evaluated using Normalized Discounted Cumulative Gain at rank 10 (NDCG@10) on DisastIR-Test. \textsc{DMRetriever} training setting is detailed in Appendix \ref{appendix:DM_train}.

\begin{figure*}[t] 
    \centering
    \includegraphics[width=\textwidth]{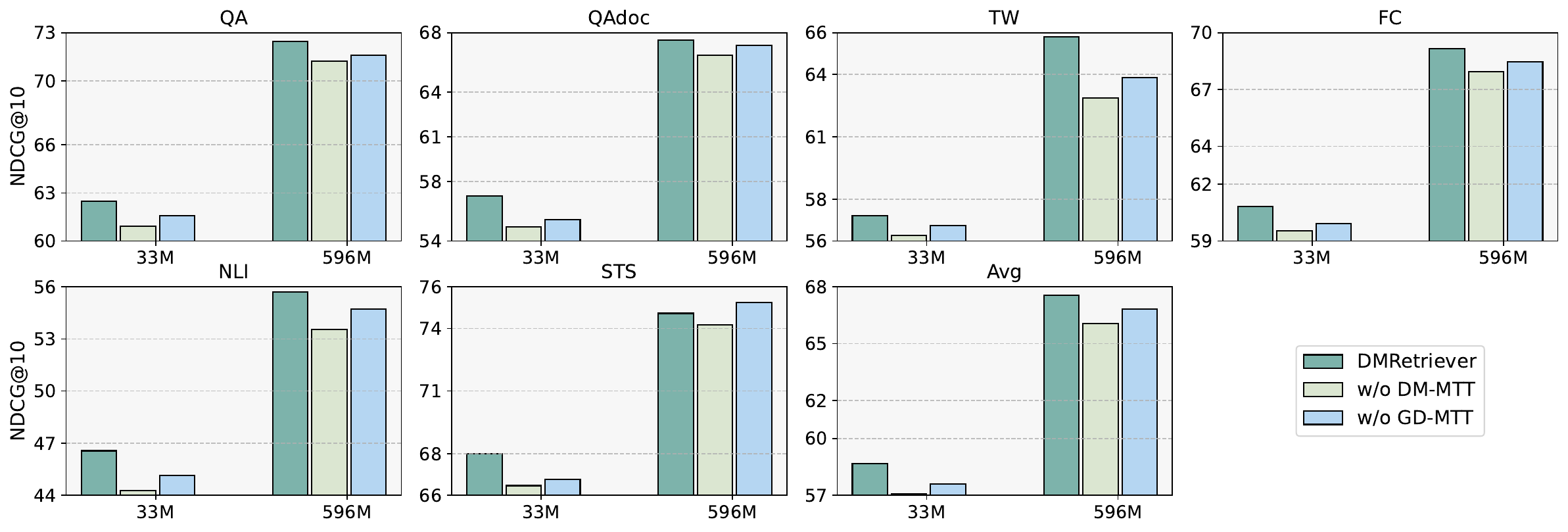}
    \caption{Effects of different fine-tuning datasets. DM-MTT and GD-MTT are datasets used during fine-tuning from disaster management and general domains. Effects of different pre-training datasets are shown in Figure \ref{fig:PT_ablation}.}
    \label{fig:FT_ablation}
\vspace{-5pt}
\end{figure*}

\section{Experiment Results and Analyses}
We investigate three research questions: \textbf{RQ1:} How does \textsc{DMRetriever} perform across scales compared to existing baselines? \textbf{RQ2:} Can our data refinement pipeline generate high-quality data for model learning? \textbf{RQ3:} Can our three-stage training framework enable models to effectively absorb intensive domain knowledge?
\subsection{\textsc{DMRetriever} Performance (RQ1)}

Table~\ref{tab:embedding_results} details \textsc{DMRetriever} performance against all baselines across all six search intents.
Across all scales, our model ranks first on every individual intent and surpasses the previous SOTA by up to 7 points in average performance (335M variant).
This performance culminates in our largest model, \textsc{DMRetriever-7.6B}, which sets a new SOTA on the benchmark with an average score of 71.37, surpassing the previous best by 4.3 points.

A key advantage of \textsc{DMRetriever} is its high efficiency. The trade-off between performance and model size is noteworthy: 
The 33M, 109M, 335M, 596M, and 4B variants retain 82.2\%, 84.0\%, 88.1\%, 94.6\%, and 98.7\% of the performance of the 7.6B model, while using only 0.4\%, 1.4\%, 4.4\%, 7.8\%, and 52.6\% of its parameters. 
This parameter efficiency leads to superior cross-scale performance. For instance, \textsc{DMRetriever-596M} surpasses all XL baselines ($\geq$4B) despite being over 13.3$\times$ smaller, and our smallest model, \textsc{DMRetriever-33M}, outperforms all baselines in medium size range using only 7.6 \% parameters.

\subsection{\textsc{DMRetriever-PT} Performance (RQ1)}

We further evaluate \textsc{DMRetriever}-PT (\textsc{DMRetriever}-Pre-Training), which uses only unlabeled data for pre-training (\S\ref{sec: PT}). As shown in Table~\ref{tab:unsup_models}, it outperforms many supervised models reported in Table~\ref{tab:embedding_results}. Notably, \textsc{4B-PT} surpasses all supervised baselines, while \textsc{596M-PT} achieves comparable performance.

This strong unsupervised performance is especially valuable in disaster management, where labeled data is costly and scarce, making our approach a practical and data-efficient way to rapidly develop high-quality IR models for this domain.

\subsection{Ablations on Data Refinement Pipeline (RQ2)}

\paragraph{Ablations on Training Dataset Component}

Figures \ref{fig:FT_ablation} and \ref{fig:PT_ablation} highlight the importance of disaster management-specific and general-domain corpora in shaping model performance. General-domain data introduces diverse language patterns and semantic relevance, enhancing model robustness and adaptability \citep{xu2024simrag}.

Disaster management corpus grounds model in the context of this domain, enhancing its ability to adapt to domain-specific scenarios. 

This yields substantial improvements, particularly on search intents that demand intensive domain knowledge (e.g., NLI and Twitter) \citep{yin2025disastir}, with gains of up to 2.9 points on Twitter in 596M variant. These results underscore the importance of disaster management-specific data for effective domain adaptation and model performance.

\begin{figure}[ht] 
    \centering
    \includegraphics[width=\columnwidth]{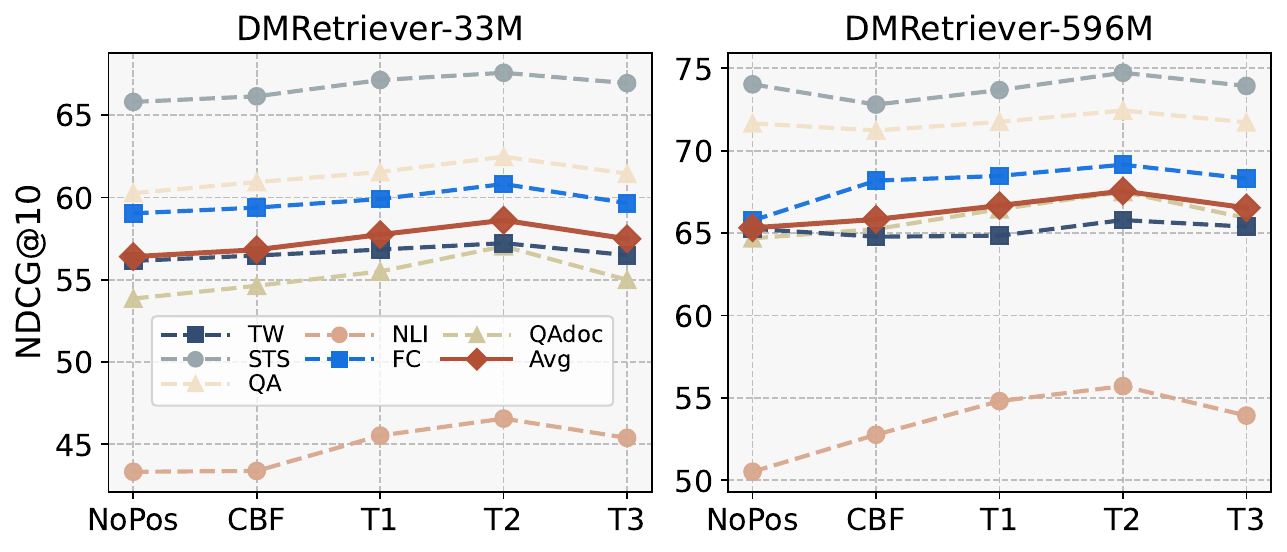}
    \caption{Ablation study of different false positive filtering methods. NoPos: No false positive filtering. CBF: Consistency-based filtering. T1--3: Top-$N$ mutual-agreement filtering strategy, where $N=1, 2, 3$.}
    \label{fig:ablation_pos_filter}
\vspace{-4pt}
\end{figure}

\begin{figure}[ht]  
    \centering
    \includegraphics[width=\columnwidth]{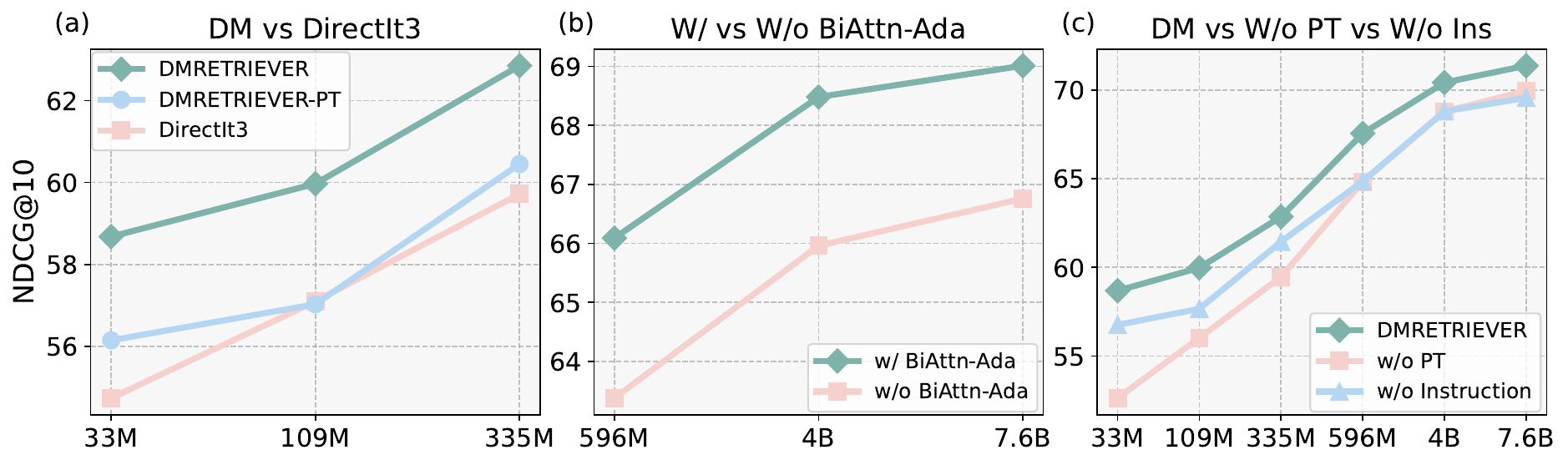}
    \caption{Ablations of progressive instruction fine-tuning. DirectIt3, BiAttn-Ada, DM-PT, and Ins mean fine-tuning directly using iteration 3 data (MTT-0.85), Bidirectional attention adaptation, \textsc{DMRetriever} model, and instruction, respectively. Breakdown for each search intent is in Figures \ref{fig:Ablation_Stage_eachtask_FT}, \ref{fig:Ablation_Stage_eachtask_PTS1}, and \ref{fig:Ablation_Stage_eachtask_PT_vs_noPT}.} 
    \label{fig:train_stage_p2}
\vspace{-6pt}
\end{figure}

\paragraph{Ablations on False Positive Filter Methods}
\label{sec: ablation_false_filter}

We further ablate our data refinement pipeline to assess its role in improving data quality and model performance. We begin with false positive filter, comparing (1) no filtering (NoPos) and (2) a consistency-based filter (CBF; details in Appendix~\ref{appendix: cbf}). Difficulty-aware hard negative mining is analyzed together with three-stage training framework in \S\ref{sec: ablation_train_stage}.

As shown in Figure~\ref{fig:ablation_pos_filter}, CBF method yields only marginal gains over NoPos, improving average performance by 0.4 and 0.5 points for 33M and 596M variants. In contrast, our top-$N$ mutual-agreement filtering consistently outperforms both baselines across most search intents and in average performance, achieving up to 2.2 points higher average performance. This shows CBF struggles to filter low-quality pairs in disaster management, mainly due to the weaknesses of individual IR models in this domain, while our mutual-agreement approach addresses this limitation through model consensus.

An appropriate choice of $N$ further enhances performance: a less strict top-$1$ agreement ($N=1$) may retain noisy pairs, while a stricter top-3 agreement ($N=3$) may exclude true positives, slightly affecting performance.

\subsection{Ablations on Three-stage Training Frameworks (RQ3)}
\label{sec: ablation_train_stage}

\paragraph{Ablations on Progressive Instruction Fine-tuning}
As shown in Figure~\ref{fig:train_stage_p2}a, progressive fine-tuning is essential for model performance. Skipping intermediate stages and directly fine-tuning \textsc{DMRetriever-PT} on the final stage dataset (details in Appendix~\ref{appendix: ablation_three_stage}) causes a sharp performance drop, with up to 1.4 points (33M variant) lower than its pre-trained variant. This degradation indicates a mismatch between model capacity and data difficulty, as the final-stage negatives are too challenging for the model’s early-stage learning.
% and degrading performance. 

In contrast, our progressive method, which gradually increases the difficulty of hard negatives, enables steady improvement across iterations (Figure~\ref{fig:performance_stage}). This confirms the \emph{necessity of progressive fine-tuning} and highlights \emph{the difficulty-aware hard negative mining} that enables progressive training by providing controlled negatives.

We also find that adding intent-specific instructions to query significantly improves performance (Figure~\ref{fig:train_stage_p2}c), with gains of up to 2.3 points for 109M variant. This suggests that instruction tuning helps model better align its embeddings with diverse search intents \citep{asai2022task, wang2023improving}.

\paragraph{Ablations on Pre-training}

Adaptation to bidirectional attention is vital for decoder-only backbones, yielding a significant performance improvement  (Figure~\ref{fig:train_stage_p2}b) with up to 2.7 points improvement (596M variant). This indicates that bidirectional attention could yield more effective embeddings than causal attention, as it allows model to capture knowledge from future tokens \citep{lee2024nv}.

Contrastive pre-training is crucial for effective domain knowledge adaptation, especially for smaller models, yielding robust initialization for retrieval tasks (Figure~\ref{fig:train_stage_p2}c). The performance gains, however, decrease with increasing model size, as larger models may already possess the capacity to capture domain knowledge \citep{xu2024bmretriever}.

\section{Conclusion}

We present \textsc{DMRetriever}, a family of six dense retrieval models (33M to 7.6B) designed to advance information access in disaster management. \textsc{DMRetriever} is trained through a novel three-stage training framework with training data produced by an advanced data refinement method. 

Extensive experiments demonstrate that \textsc{DMRetriever} sets a new SOTA across all six search intents within each model scale. Furthermore, \textsc{DMRetriever} shows exceptional parameter efficiency, with the 596M version outperforming all XL baselines ($\geq$4B) over 13.3$\times$ its size, and the 33M version surpassing all medium-sized competitors using only 7.6\% of their parameters.

\section*{Limitations}
While \textsc{DMRetriever} achieves SOTA performance across all search intents and scales, several aspects merit further investigation. The current work focuses primarily on textual retrieval; extending it to multi-modal retrieval would enhance its applicability in real-world disaster scenarios. In addition, \textsc{DMRetriever} is focused on English-language resources, and future work could explore multilingual and cross-lingual retrieval to broaden its global utility.

\section*{Ethics Statement}
\textsc{DMRetriever} is designed to support disaster management by enhancing retrieval performance across diverse search intents. All training data are drawn from publicly available sources, with no personally identifiable information included. We acknowledge potential risks of misuse, such as the amplification of rumors during disasters. To mitigate such concerns, \textsc{DMRetriever} is released strictly for research purposes.

A potential issue is test set contamination, where some test samples may overlap with training data \citep{sainz2023nlp}. We conduct detailed data contamination analysis in Appendix \ref{appendix:data_contain}, where we do not observe any overlap between train and test sets.

\section*{Acknowledgments}
This work used DeltaAI and Delta GPU at the National Center for Supercomputing Applications through allocation CIV250019, CIS250751, and CIV250021 from the Advanced Cyberinfrastructure Coordination Ecosystem: Services \& Support (ACCESS) program, which is supported by U.S. National Science Foundation grants \#2138259, \#2138286, \#2138307, \#2137603, and \#2138296.

\bibliography{custom}

\newpage

\appendix

\begin{algorithm}[t]
% \SetAlgoNoLine
\caption{Three-stage training pipeline of \textsc{DMRetriever}}
\label{alg:dmretriever_train}
\KwIn{ \\
  Backbone $\mathcal{M}$ (encoder- or decoder-only); 
  MLM training data $\mathcal{D}_{MLM}$; 
  Difficulty level schedule $\{\alpha_t\}$ for progressive fine-tuning
}
\KwOut{Final \textsc{DMRetriever} model}
\textbf{Stage 1: Bidirectional attention adaptation (decoder-only backbone)} \\
\If{$\mathcal{M}$ is decoder-only}{
  Replace causal mask with full attention mask; \\
  Train $\mathcal{M}$ on $\mathcal{D}_{MLM}$ using the MLM objective
}
\textbf{Stage 2: Unsupervised contrastive pre-training} \\
Construct $\mathrm{MTP}$ via Algorithm~\ref{alg:dmrpt}; \\
Pre-train $\mathcal{M}$ on $\mathrm{MTP}$ using InfoNCE loss (Eq.~\ref{eq:cpt_infonce}) with in-batch negatives \\
\textbf{Stage 3: Difficulty-aware Progressive Supervised Instruction Fine-tuning} \\
\ForEach{iteration $t$}{
  Construct $\mathrm{MTT}\text{-}\alpha_t$ via Algorithm~\ref{alg:dmrft} with difficulty level $\alpha_t$; \\
  Fine-tune $\mathcal{M}$ on $\mathrm{MTT}\text{-}\alpha_t$ using supervised InfoNCE loss (Eq.~\ref{eq:hnm_infonce}); \\
  Update $\mathcal{M}$ with the best validation checkpoint
}
\Return $\mathcal{M}$
\end{algorithm}
% \vspace{-5pt}

\begin{algorithm}[t]
\caption{Construction of MTP}
\label{alg:dmrpt}
\KwIn{ \\
  $\mathcal{C}_{GD}$: general-domain corpora; 
  $\mathcal{R}_{dm}$: raw disaster-management documents (e.g., PDFs); 
  $\mathcal{I}$: search intents; 
  $\Gamma_i$: heuristic rule for intent $i$; 
  $\mathcal{E}^{need}_i, \mathcal{E}^{query}_i$: in-context examples for information-need generation, and for query--positive passage generation; \\
  $\mathrm{Prompt}^{need}, \mathrm{Prompt}^{query}$: prompt templates for information-need generation, and for query--positive passage generation
}
\KwOut{ \\
  $\mathrm{MTP} = \mathrm{GD\mbox{-}MTP} 
  \cup \mathrm{DM\mbox{-}MTP}$ \\
} 

\textbf{1. Build DM-MTP:} \\
\ForEach{$d \in \mathcal{R}_{dm}$}{
  $\mathcal{C}_{dm} \gets \mathrm{ProcessToPassages}(d)$; \\
}
\ForEach{$p \in \mathcal{C}_{dm}$}{
  $s \gets \mathrm{LLM}(p,\mathcal{I},\mathcal{E}^{need}_i,\mathrm{Prompt}^{need})$; \\
  $(q,p^+) \gets \mathrm{LLM}(s,p,\mathcal{E}^{query}_i,\mathrm{Prompt}^{query})$; \\
  $\mathrm{DM\mbox{-}MTP} \gets \mathrm{DM\mbox{-}MTP} \cup \{(q,p^+)\}$; \\
}

\textbf{2. Build GD-MTP:} \\
\ForEach{$r \in \mathcal{C}_{GD}$}{
  Identify intent $i$ (by dataset design); \\
  $(q,p^+) \gets \mathrm{ApplyHeuristic}(r,\Gamma_i)$; \\
  $\mathrm{GD\mbox{-}MTP} \gets \mathrm{GD\mbox{-}MTP} \cup \{(q,p^+)\}$; \\
}

\textbf{3. Combine and output:} \\
$\mathrm{MTP} \gets \mathrm{GD\mbox{-}MTP} 
\cup \mathrm{DM\mbox{-}MTP}$
\end{algorithm}

\begin{table*}[t]
\centering
\small
\setlength{\tabcolsep}{4pt} 
\begin{tabular}{lcccccc}
\toprule
\textbf{} & \textbf{QA} & \textbf{QAdoc} & \textbf{Twitter} & \textbf{FactCheck} & \textbf{NLI} & \textbf{STS} \\
\midrule
Top 1 & NV-Embed-v2 & Linq-Emb-Mis & SFR-Emb-Mis & Linq-Emb-Mis & NV-Embed-v2 & NV-Embed-v2 \\
Top 2 & Linq-Emb-Mis & NV-Embed-v2 & inf-retriever-v1 & SFR-Emb-Mis & gte-Qwen2-1.5B-ins & inf-retriever-v1 \\
Top 3 & inf-retriever-v1 & SFR-Emb-Mis & Linq-Emb-Mis & NV-Embed-v2 & inf-retriever-v1-1.5b & gte-Qwen2-7B-ins \\
\bottomrule
\end{tabular}
\caption{Top-3 IR models across different search intents identified in DisastIR. Linq-Emb-Mis and SFR-Emb-Mis are shorten for Linq-Embed-Mistral and SFR-Embedding-Mistral}
\label{tab:ir-top-models}
\end{table*}

\begin{table*}[ht]
\centering
\small
\renewcommand{\arraystretch}{1.1}
\begin{tabular*}{\textwidth}{@{\extracolsep{\fill}} l p{0.8\textwidth}}
\hline
\textbf{Task} & \textbf{Instruction Format} \\
\hline
QA & Given the question, retrieve most relevant passage that best answers the question \\

QAdoc & Given the question, retrieve most relevant document that answers the question \\

Twitter & Given the user query, retrieve the most relevant Twitter text that meets the request \\

FactCheck & Given the claim, retrieve most relevant document that supports or refutes the claim \\

NLI & Given the premise, retrieve most relevant hypothesis that is entailed by the premise \\

STS & Given the sentence, retrieve the sentence with the same meaning \\

\hline

\end{tabular*}
\caption{Instruction formats used for each search intent.} 
\vspace{-5pt}
\label{tab:PT_instruction_formats}
\end{table*}

\begin{algorithm}[t]
\caption{Construction of MTT-$\alpha$}
\label{alg:dmrft}
\KwIn{ \\
  $\mathcal{Q}_{GD}$: queries from GD-MTP (with $(q,p^+)$ pairs); 
  $\mathcal{Q}_{DM}$: queries from DM-MTP; 
  $N$: top-$N$ passages for mutual-agreement; 
  $\mathcal{M}_{top3}$, $\mathcal{M}_{best}$: top-3 and best IR models per intent (from DisastIR); 
  $\alpha$: difficulty level for hard negative selection
}
\KwOut{ $\mathrm{MTT}\text{-}\alpha = \mathrm{GD\mbox{-}MTT}\text{-}\alpha \cup \mathrm{DM\mbox{-}MTT}\text{-}\alpha$ }

\textbf{1. Mutual-agreement false positive filter} \\
\ForEach{$q \in \mathcal{Q}_{DM}$}{
  Identify intent $i$; \\
  \ForEach{$m \in \mathcal{M}_{top3}(i)$}{
    Retrieve top-$N$ passages $R_m(q)$; \\
  }
  $\mathcal{C}_q \gets \bigcap_{m \in \mathcal{M}_{top3}(i)} R_m(q)$; \\
  \If{$\mathcal{C}_q \neq \emptyset$}{
    Add $(q, \text{top-1}(\mathcal{C}_q))$ to $\mathcal{D}^+_{DM}$; \\
  }
}
$\mathcal{D}^+ \gets \mathcal{D}^+_{DM} \cup \mathcal{Q}_{GD}$; \\

\textbf{2. Difficulty-aware hard negative mining} \\
\ForEach{$(q,p^+) \in \mathcal{D}^+$}{
  Identify intent $i$; \\
  Retrieve top-200 $R_{best}(q)$ using $\mathcal{M}_{best}(i)$; \\
  $\mathcal{H}_q \gets \{p^- \in R_{best}(q)\setminus\{p^+\} \mid 
   \mathrm{sim}(q,p^-) < \mathrm{sim}(q,p^+) \times \alpha \}$; \\
  Add $(q,p^+,\text{top-}k(\mathcal{H}_q))$ to 
  $\mathrm{DM\mbox{-}MTT}\text{-}\alpha$ if $q\in\mathcal{Q}_{DM}$ else $\mathrm{GD\mbox{-}MTT}\text{-}\alpha$; \\
}

\textbf{3. Combine and output:} 
$\mathrm{MTT}\text{-}\alpha \gets \mathrm{GD\mbox{-}MTT}\text{-}\alpha \cup \mathrm{DM\mbox{-}MTT}\text{-}\alpha$; \\
\Return $\mathrm{MTT}\text{-}\alpha$
\end{algorithm}

\section{Details of DM-MTP Construction}
\label{appendix:DM-MTP}
We propose a three-stage pipeline to construct DM-MTP: (1) web crawling and processing of disaster-management domain PDF files, (2) information needs generation, and (3) query–positive passage pairs generation. 

For web crawling, we construct queries in the format “event type name + location name + .pdf”. We specifically target PDFs because they are more likely to contain structured, information-rich, and credible content, often originating from peer-reviewed publications or official institutions. To ensure coverage, we consider 301 disaster event types as defined in \citet{undrr2020hazard} and combine them with the names of the 50 U.S. states, given both the availability of high-quality resources in the U.S. context and the fact that our model is designed for English-language data. 

All collected PDF files are then processed into semantically coherent text chunks through a multi-step pipeline proposed by \citet{yin2025disastir}: (1) exact-URL deduplication, where duplicate documents are removed by identifying identical download links; (2) text extraction and preprocessing, where each PDF is converted into plain text and non-textual elements such as tables and figures are removed; (3) locality-sensitive hashing (LSH) deduplication, which eliminates near-duplicate documents with overlapping content; (4) semantic chunking, where cleaned documents are segmented into passages of fewer than 256 tokens to balance retrievability and semantic integrity; and (5) embedding-based near deduplication, where dense embeddings are computed for all chunks, an ANN index is built, and pairs with cosine similarity above 0.9 are discarded. Chunks obtained are regarded as disaster management passages in DM-MTP.

In the second and third stages, given a passage and the definition of a search intent, an LLM generates information-need statement. The information statement together with the passage is used to construct a user query and its directly relevant positive passage. Prompts for query generation and relevant passage generation based on disaster management-related passages under different search intents are adapted from \citet{yin2025disastir}. 

In total, DM-MTP contains 3.3 million query–passage pairs, with an estimated generation cost of about \$2,000 using the GPT-4o-mini API.

\section{Details of GD-MTP Construction}
\label{appendix:PT_set}

To construct DG-MTP, we generate query–positive passage pairs from publicly available unlabeled corpora using intent-specific strategies as follows:
(1) For QA, QAdoc, and TW intents, the user question serves as the query, and its corresponding answer is the positive passage.
(2) For FC intent, the claim is used as the query, and a document that supports or refutes it is selected as the positive passage.
(3) For NLI intent, the premise is treated as the query and the entailed hypothesis as the positive passage.
(4) For STS intent, one sentence from a similar pair is randomly chosen as the query and the other as the positive passage.

\section{Data Contamination Analysis}
\label{appendix:data_contain}
Following \citet{wang2023improving}, we conduct test set data contamination analysis from two aspects: query overlap and passage corpus overlap.

\paragraph{Query overlap.} Following \citet{wang2023improving}, we perform a string-matching analysis between queries in the DisastIR-Test and those in the training sets (MTP and MTT), and find no overlap between training and test queries.

\paragraph{Passage corpus overlap.} While sharing the same passage corpus (e.g., DBPedia, NQ, and TriviaQA all using Wikipedia) is a standard evaluation practice in information retrieval and is not regarded as contamination \citep{wang2023improving}, we adopt a stricter criterion by performing a string-matching analysis between passages in DisastIR-Test and those in MTP and MTT, finding no overlap between the training and test corpora.

\section{Determination of $\alpha$ Value during Progressive Fine-tuning}
\label{appendix:alpha_choice}

We follow the ablation study results of \citet{moreira2024nv} and set $\alpha=0.95$ for \textsc{DMRetriever-596M}, \textsc{4B}, and \textsc{7.6B}.

For \textsc{DMRetriever-33M}, \textsc{109M}, and \textsc{335M}, we apply progressive fine-tuning across three iterations, training each model for one epoch per iteration. We consider three candidate ranges of $\alpha$: $\{0.55, 0.65, 0.75\}$, $\{0.65, 0.75, 0.85\}$, and $\{0.75, 0.85, 0.95\}$. Model performance is assessed on the validation set (DisastIR-DevLite), and the range yielding the best validation performance is selected. Results show that models progressively trained on MTT-0.65, 0.75, and 0.85 could lead to the best validation performance and $\{0.65, 0.75, 0.85\}$ is thus selected.

\begin{algorithm}[ht]
\caption{Construction of DisastIR-DevLite and DisastIR-Test}
\label{alg:disastir_devlite}
\KwIn{ \\
  DisastIR corpus $\mathcal{C}$ with labeled queries $\mathcal{Q}$ and relevance $\mathcal{Y}$; Search intents $\mathcal{T}$; Top-3 IR models $\mathcal{M}_{top3}$ per intent; $N=80$ queries per intent for DevLite, $k=10$ passages per model
}
\KwOut{ \\
  DisastIR-DevLite $(\mathcal{Q}_\text{dev}, \mathcal{C}_\text{dev}, \mathcal{Y}_\text{dev})$; \\
  DisastIR-Test $(\mathcal{Q}_\text{test}, \mathcal{C}, \mathcal{Y}_\text{test})$
}

\textbf{Step 1: Sample queries for DevLite and Test} \\
\ForEach{intent $t \in \mathcal{T}$}{
  Randomly sample $N$ queries $\mathcal{Q}_\text{dev}^{(t)} \subset \mathcal{Q}^{(t)}$; \\
  Let $\mathcal{Q}_\text{test}^{(t)} \gets \mathcal{Q}^{(t)} \setminus \mathcal{Q}_\text{dev}^{(t)}$
}
$\mathcal{Q}_\text{dev} \gets \bigcup_t \mathcal{Q}_\text{dev}^{(t)}$, \quad
$\mathcal{Q}_\text{test} \gets \bigcup_t \mathcal{Q}_\text{test}^{(t)}$ \\[2pt]

\textbf{Step 2: Build passage pool per DevLite query} \\
\ForEach{$q \in \mathcal{Q}_\text{dev}$ with intent $t$}{
  \ForEach{$m \in \mathcal{M}_{top3}(t)$}{
    Retrieve $\text{TopK}_m(q)$ from $\mathcal{C}$
  }
  $\text{Pool}(q) \gets \bigcup_{m \in \mathcal{M}_{top3}(t)} \text{TopK}_m(q)$
}
$\mathcal{C}_\text{dev} \gets \bigcup_{q \in \mathcal{Q}_\text{dev}} \text{Pool}(q)$ \\[2pt]

\textbf{Step 3: Inherit relevance labels from DisastIR} \\
$\mathcal{Y}_\text{dev} \gets \mathcal{Y}|_{\mathcal{Q}_\text{dev} \times \mathcal{C}_\text{dev}}$; \quad
$\mathcal{Y}_\text{test} \gets \mathcal{Y}|_{\mathcal{Q}_\text{test} \times \mathcal{C}}$ \\

\Return DisastIR-DevLite and DisastIR-Test
\end{algorithm}

\section{Details of DisastIR-DevLite and DisastIR-Test Construction}
\label{appendix:disastir_devlite}
To facilitate efficient yet effective model development, we construct a lightweight validation set, referred to as DisastIR-DevLite (See Algorithm \ref{alg:disastir_devlite}). For each search intent, we randomly sample 80 user queries from DisastIR to form DisastIR-DevLite, while the remaining queries constitute DisastIR-Test. For each query in DisastIR-DevLite, we apply the top three IR models corresponding to its search intent, retrieving the top-10 passages from each model. The union of these top-10 passages forms a passage pool for that query. The final DisastIR-DevLite passage corpus is constructed as the union of the passage pools of all queries in DisastIR-DevLite, whereas DisastIR-Test passage corpus is the entire DisastIR passage corpus. Query–passage relevance labels for both DisastIR-DevLite and DisastIR-Test are directly inherited from the original DisastIR. 

\section{Efficiency and Effectiveness of DisastIR-DevLite}
\label{appendix:disastir_dev}

To evaluate the efficiency and effectiveness of DisastIR-DevLite, we compare it against DisastIR-DevFull, a counterpart that shares the same queries as DisastIR-DevLite but uses the complete DisastIR corpus. For our analysis, we sampled 100 checkpoints from each model, \textsc{DMRetriever-33M} and \textsc{DMRetriever-596M}, during training, and evaluated each checkpoint on both development sets. The evaluation metrics include average embedding time, as well as average and intent-level NDCG@10 score.

\begin{figure}[htbp]
    \centering
     \includegraphics[width=\linewidth]{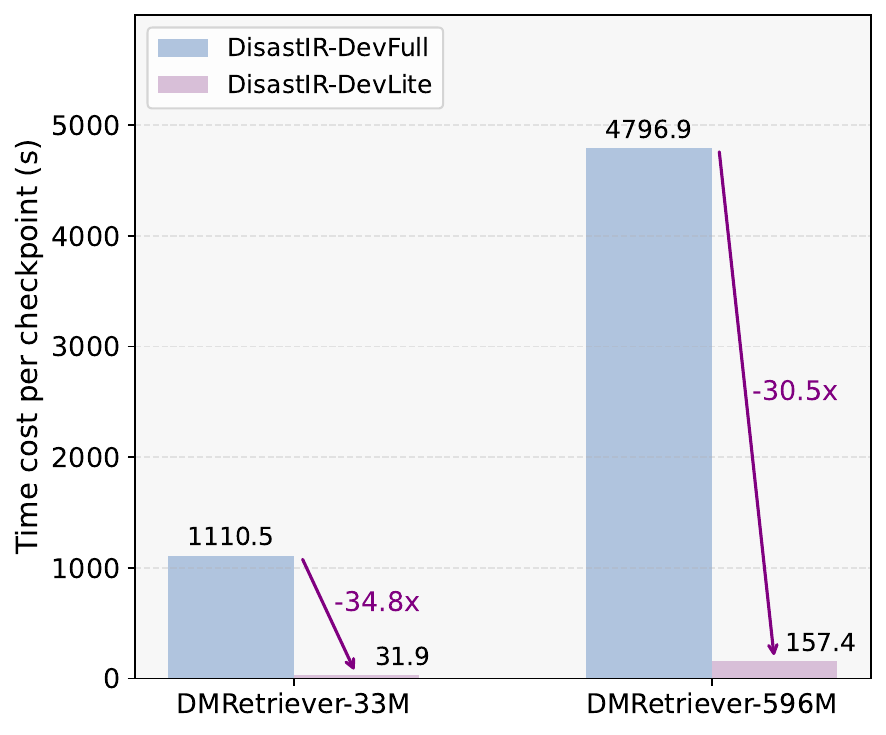}
    \caption{Efficiency comparison between DisastIR-DevLite and DisastIR-DevFull.}
    \label{Fig: time_dev}
% \vspace{-11pt}
\end{figure}

The results demonstrate substantial efficiency gains. As shown in Figure~\ref{Fig: time_dev}, DisastIR-DevLite is 34.8$\times$ and 30.5$\times$ faster than DisastIR-DevFull for \textsc{DMRetriever-33M} and \textsc{DMRetriever-596M}, respectively. Crucially, this speedup does not compromise effectiveness. The checkpoint performance rankings produced by DisastIR-DevLite are nearly identical to those from DisastIR-DevFull, achieving a Kendall’s $\tau$ correlation of over 0.90 for both models (Table~\ref{tab:Kendall_dev}).

\begin{table}[th]
\centering
\begin{tabular}{lcc}
\hline
Dataset & \#Queries & \#Passages \\
\hline
DisastIR-DevLite & 480  & 9,316   \\
DisastIR-Test    & 9,120 & 239,704 \\
\hline
\end{tabular}
\caption{Statistics of DisastIR-DevLite and DisastIR-Test.}
\label{tab:disastir-stats}
\end{table}

\section{Detailed Reasons for Excluding Knowledge Distilled (KD) Baselines}
\label{appendix:remove_kd}

For baselines \citep{wang2022text, awasthy2025granite} that use knowledge distillation (KD) during fine-tuning, a cross-encoder is typically employed as the teacher model.
This process introduces substantial external knowledge, as the teacher cross-encoder has a more powerful architecture. Unlike bi-encoders (our primary baselines and \textsc{DMRetriever}), cross-encoders process queries and documents jointly, allowing deep token-level interactions that produce highly accurate relevance scores. These fine-grained scores are then used as soft labels for the student model, offering a much richer and more informative training signal than the sparse binary hard labels available to other models. 

Consequently, student model does not learn from ground-truth data alone, but is guided by teacher’s high-quality interpretation of that data.

For this reason, we exclude all KD-based baselines, which include E5-small, base, large-v2 \citep{wang2022text}, and granite-embedding-125m-english \citep{awasthy2025granite}, to ensure fair comparisons.

\section{Information on Baseline Models and Their Implementation}
\label{appendix:eva_models}

We adopt baseline models evaluated in DisastIR and additionally include recent models from MTEB that were not evaluated in DisastIR. They are categorized as four scales based on model parameter sizes: small (<=109M), medium (137M - 335M), large (434M- 1.5B), and extra large (XL) (>= 4B). 

Detailed specifications of the selected models are provided in Table~\ref{tab:model_specs}, while their HuggingFace links and licenses are listed in Table~\ref{tab:model_links}. 

For each baseline model, we adhere to the official implementation guidelines to generate normalized query and passage embeddings. All evaluations are conducted in a zero-shot setting, with input sequences truncated to 512 tokens and a task-specific instruction prepended to each query.

\section{Implementation Details of \textsc{DMRetriever} Training}
\label{appendix:DM_train}

Backbones of \textsc{DMRetriever} are shown in Table \ref{tab:DMretriever_models}. For bidirectional attention adaptation of decoder-only backbones, we train for 2 epochs with a global batch size of 256. The learning rates are set according to model size (see Table~\ref{tab:training-hyperparams}). All experiments are conducted on a single H200 GPU.  

For unsupervised contrastive pre-training, each model is trained for 1 epoch. We use 3 H200 GPUs for encoder-only backbones and 4 H200 GPUs for decoder-only backbones. To optimize GPU memory consumption, we adopt DeepSpeed ZeRO Stage 1 \citep{rasley2020deepspeed} and gradient checkpointing. Gradient accumulation is set to 1, and the per-GPU batch size is maximized to increase the diversity of in-batch negatives \citep{bge_embedding}, leading to the global batch sizes reported in Table~\ref{tab:training-hyperparams}.  

For progressive supervised fine-tuning, we train encoder-only backbones for 1 epoch per iteration and decoder-only backbones for 1 epoch. Fine-tuning is conducted on a single H200 GPU, with gradient checkpointing disabled to improve training speed and gradient accumulation enabled. The learning rates and global batch sizes for each model variant are listed in Table~\ref{tab:training-hyperparams}.  

Across all experiments, we use brain floating point (bfloat16) quantization and truncate input sequences to 512 tokens. The temperature value is set as 0.01. For each query, we set one positive passage and nine passages during pre-training and fine-tuning. For decoder-only backbones, we apply LoRA with rank $r=16$ and $\alpha=32$, following prior work \citep{lee2024nv, wang2023improving, xu2024bmretriever}, leaving hyperparameter tuning for future exploration.  

\begin{table*}[t]
\centering
\small
\begin{tabular}{lcccccc}
\toprule
\textbf{} & \textbf{33M} & \textbf{109M} & \textbf{335M} & \textbf{596M} & \textbf{4B} & \textbf{7.6B} \\
\midrule
\textbf{Bidirectional Adaptation Stage} & & & & & & \\
\cmidrule(lr){1-7}
Learning Rate & - & - & - & 1e-4 & 5e-5 & 2e-5 \\
Batch Size    & - & - & - & 256  & 256  & 256  \\
\midrule
\textbf{Contrastive Pre-training} & & & & & & \\
\cmidrule(lr){1-7}
Learning Rate & 4e-4 & 3e-4 & 1e-4 & 9e-5 & 4e-5 & 2e-5 \\
Batch Size    & 2400 & 732  & 384  & 384  & 192  & 140  \\
\midrule
\textbf{Supervised Fine-tuning} & & & & & & \\
\cmidrule(lr){1-7}
Learning Rate & 4e-5 & 1e-5 & 9e-6 & 6e-5 & 2e-5 & 1e-5 \\
Batch Size    & 1024 & 512  & 512  & 512  & 128  & 64   \\
\bottomrule
\end{tabular}
\caption{Training hyperparameters for different model sizes during the three-stage training framework.}
\label{tab:training-hyperparams}
\end{table*}

\section{Implementation of Consistency-based filtering (CBF) Method}
\label{appendix: cbf}
To implement the CBF method, we follow the methodology of \citet{wang2022text}. Specifically, for each search intent, we use its best-performing retrieval model identified from the DisastIR benchmark as a reference (Table \ref{tab:ir-top-models} for best-performing retrieval models). A synthetic pair is then discarded if its LLM-generated passage is not ranked within the top-2 results retrieved by this reference model.

\section{Implementation of Ablations of Progressive Fine-tuning}
\label{appendix: ablation_three_stage}

In the ablation study of progressive fine-tuning, we fine-tune \textsc{DMRetriever-33M-PT}, \textsc{109M-PT}, and \textsc{335M-PT} on MTT-0.85 for 3 epochs. All other hyperparameters, including learning rate and batch size, are kept the same as in progressive fine-tuning which are detailed in Appendix \ref{appendix:DM_train}.

\section{Performance of \textsc{DMRetriever-KD}}

To establish a fair comparison with baselines like the E5 series that employ knowledge distillation (KD) during model fine-tuning and boost performance of our smaller models, we apply KD to our fine-tuned \textsc{DMRetriever-33M, 109M, and 335M} models. This process yields \textsc{DMRetriever-KD} series (see Appendix~\ref{appendix:dm_kd} for KD settings). As shown in Table~\ref{tab:dm_kd}, our \textsc{DMRetriever-KD} achieves the best average performance within their respective scales, securing the top rank across the majority of search intents.

The parameter efficiency of these models is remarkable. Notably, \textsc{DMRetriever-33M-KD} surpasses all 560M-series baselines despite being 17$\times$ smaller. Furthermore, \textsc{DMRetriever-335M-KD} matches models in the 7B-parameter class, which are over 20$\times$ larger. This makes our KD-series models ideal for real-world disaster management applications, where computational resources are limited and low latency is critical.

\section{Knowledge Distillation Settings}
\label{appendix:dm_kd}

We apply knowledge distillation (KD) to our \textsc{DMRetriever-33M, 109M, and 335M} models, creating their distilled counterparts, denoted as the \textsc{DMRetriever-KD} series. The distillation is performed using a specially constructed dataset, MTT-0.95-KD.

This dataset is an extension of MTT-0.95, the construction of which is detailed in Section~\ref{section: train_S3} and Algorithm~\ref{alg:dmrft} (Lines 10--16). To create KD version, we use the best-performing IR model, as identified in DisastIR, for each search intent as a teacher to generate query-passage similarity scores. These scores are then added to the dataset as soft labels.

The training objective for knowledge distillation, $\mathcal{L}_{\text{KD}}$, is formulated as a weighted sum of two components. It combines the contrastive loss, $\mathcal{L}_{\text{CL}}$ (from Eq.~\ref{eq:hnm_infonce}), on the original hard labels with a Kullback-Leibler (KL) divergence term, $\mathcal{L}_{\text{KL}}$, for distilling the soft labels from the teacher. The balance between these two is controlled by a hyperparameter $\lambda$, as shown below:
$$
\mathcal{L}_{\text{KD}} = \mathcal{L}_{\text{CL}} + \lambda \mathcal{L}_{\text{KL}}
$$
In this work, we set the weighting factor $\lambda$ to 1 for all distillation experiments.

\begin{table*}[ht]
\centering
\small
\renewcommand{\arraystretch}{1.05}
\setlength{\tabcolsep}{3pt}
\begin{tabular*}{\textwidth}{@{\extracolsep{\fill}}l|c|cccccc|c}
\hline
\textbf{Model} & \textbf{Scale} 
    & \textbf{QA} & \textbf{QAdoc} & \textbf{TW} & \textbf{FC} 
    & \textbf{NLI} & \textbf{STS} & \textbf{Avg} \\
\hline
\rowcolor{lightgray}
\multicolumn{9}{l}{\textbf{Small Size ($\leq$109M)}} \\
\hline

e5-base-V2 \citeyearpar{wang2022text} 
    & 109M 
    & 65.52 & 62.68 & 57.53 & 61.86 & 45.40 & \underline{74.38} & 61.23 \\

e5-small-V2 \citeyearpar{wang2022text} 
    & 33M  
    & 65.79 & 62.64 & 60.12 & 61.80 & 46.99 & \textbf{74.54} & 61.98 \\

\rowcolor{green!20}
\textsc{DMRetriever-33M-KD} (ours)   
    & 33M
    & \underline{68.28} & \underline{64.31} & \underline{62.92} 
    & \underline{64.88} & \underline{49.93} & 72.70 & \underline{63.84} \\

\rowcolor{green!20}
\textsc{DMRetriever-109M-KD} (ours)   
    & 109M
    & \textbf{69.61} & \textbf{66.14} & \textbf{65.55} 
    & \textbf{66.30} & \textbf{52.50} & 73.95 & \textbf{65.68} \\

\hline
\rowcolor{lightgray}
\multicolumn{9}{l}{\textbf{Medium Size (109M--434M)}} \\
\hline

granite-embedding-125m \citeyearpar{awasthy2025granite} 
    & 125M 
    & \underline{64.58} & 60.73 & 46.47 
    & \underline{62.45} & 48.05 & 71.61 & 58.98 \\

e5-large-V2    \citeyearpar{wang2022text}  
    & 335M 
    & 59.92 & \underline{63.07} & \underline{55.37} 
    & 61.85 & \underline{50.69} & \textbf{74.69} & \underline{60.93} \\

\rowcolor{green!20}
\textsc{DMRetriever-335M-KD} (ours)  
    & 335M
    & \textbf{71.86} & \textbf{67.56} & \textbf{63.56} 
    & \textbf{67.16} & \textbf{52.96} & \underline{73.74} & \textbf{66.14} \\
\hline
\end{tabular*}
\caption{Performance of knowledge distillation enhanced retrieval models across multiple search intents.}
\label{tab:dm_kd}
\vspace{-5pt}
\end{table*}

\begin{table*}[ht]
\centering
\begin{tabular}{lccccccc}
\hline
Model & QA & QAdoc & TW & FC & NLI & STS & Overall \\
\hline
\textsc{DMRetriever-33M}  & 0.8467 & 0.7504 & 0.7326 & 0.8859 & 0.8610 & 0.7255 & 0.9008 \\
\textsc{DMRetriever-596M} & 0.8995 & 0.7937 & 0.8042 & 0.9101 & 0.8730 & 0.7725 & 0.9033 \\
\hline
\end{tabular}
\caption{Effectiveness of DisastIR-DevLite. Kendall’s $\tau$ value between DisastIR-DevLite and DisastIR-DevFull.}
\label{tab:Kendall_dev}
\end{table*}

\begin{figure*}[ht] 
    \centering
    \includegraphics[width=\textwidth]{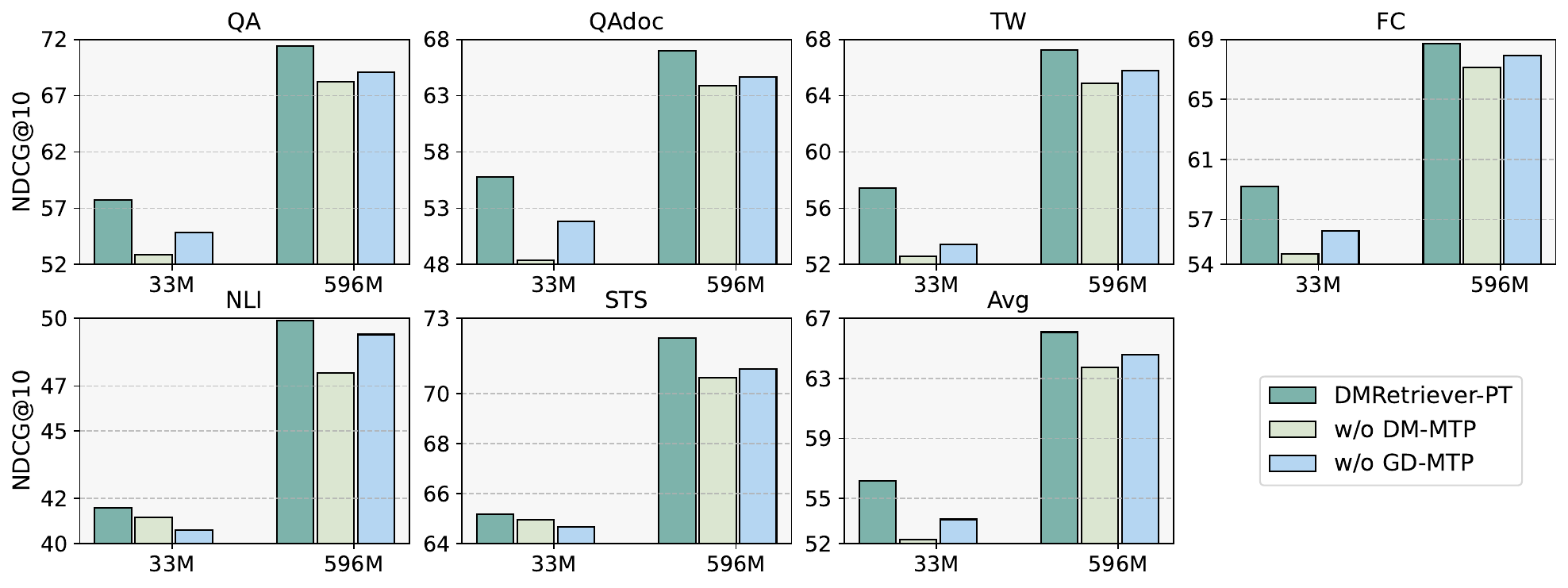}
    \caption{Effects of different pre-training datasets. DM-MTP and GD-MTP are datasets used during pre-training from disaster management and general domains.}
    \label{fig:PT_ablation}
\end{figure*}

\begin{table*}[ht]
\centering
\small
\renewcommand{\arraystretch}{1.1}
\begin{tabular*}{\textwidth}{@{\extracolsep{\fill}}l l l p{0.53\textwidth}}
\hline
\textbf{Dataset} & \textbf{Size} & \textbf{Task} & \textbf{Link} \\
\hline

\multicolumn{4}{@{}l}{\textbf{GD-MTP:}} \\
\hline

Squad & 87.6K & QA & \url{https://huggingface.co/datasets/sentence-transformers/squad} \\
Paq & 4M\textsuperscript{*} & QA & \url{https://huggingface.co/datasets/sentence-transformers/paq} \\
MSMARCO & 1M & QA & \url{https://huggingface.co/datasets/microsoft/ms_marco} \\
Gooaq & 3M\textsuperscript{*} & QA & \url{https://huggingface.co/datasets/sentence-transformers/gooaq} \\
Eli5 & 325k & QA & \url{https://huggingface.co/datasets/sentence-transformers/eli5} \\
Nfcorpus & 134K & QAdoc & \url{https://huggingface.co/datasets/mteb/nfcorpus} \\
MSMARCO-Document-Ranking & 250k\textsuperscript{*} & QAdoc & \url{https://microsoft.github.io/msmarco/Datasets} \\
Hotpotqa & 170k & QAdoc & \url{https://huggingface.co/datasets/BeIR/hotpotqa} \\
Signal1m-generated-queries & 1M\textsuperscript{*} & Twitter & \url{https://huggingface.co/datasets/BeIR/signal1m-generated-queries} \\
Fever & 250k\textsuperscript{*} & FactCheck & \url{https://huggingface.co/datasets/mteb/fever} \\
Quora & 15k & STS & \url{https://huggingface.co/datasets/mteb/quora} \\
Quora-duplicates & 950k\textsuperscript{*} & STS & \url{https://huggingface.co/datasets/sentence-transformers/quora-duplicates} \\
Altlex & 113k & STS & \url{https://huggingface.co/datasets/sentence-transformers/altlex} \\
XNLI & 133k\textsuperscript{*} & NLI & \url{https://huggingface.co/datasets/mteb/xnli} \\
All-NLI & 321k\textsuperscript{*} & NLI & \url{https://huggingface.co/datasets/sentence-transformers/all-nli} \\
\hline
\textbf{DM-MTP} & 3.29M \\
\hline

\textbf{Total} & 15M \\
\hline
\end{tabular*}
\caption{Composition of the MTP dataset used for unsupervised contrastive pre-training in Stage~2. 
% In addition to GD-MTP, MTP also includes DM-MTP, which comprises 3.29M pairs spanning all six search intent types. 
\textsuperscript{*} indicates that only a subset of the original dataset is used.}
\label{tab:PT_data}
\end{table*}

\begin{table*}[ht]
\centering
\small
\renewcommand{\arraystretch}{1.1}
\begin{tabular*}{\textwidth}{@{\extracolsep{\fill}}@{\hspace{5pt}}l l l p{0.53\textwidth}@{\hspace{5pt}}}
\hline
\textbf{Dataset} & \textbf{Size} & \textbf{Task} & \textbf{Link} \\
\hline

\multicolumn{4}{l}{\textbf{GD-MTT:}} \\
\hline

MSMARCO & 141k & QA & \url{https://huggingface.co/datasets/microsoft/ms_marco} \\

MSMARCO-Document-Ranking & 57k & QAdoc & \url{https://microsoft.github.io/msmarco/Datasets} \\

Fever & 81k & FactCheck & \url{https://huggingface.co/datasets/mteb/fever} \\

Quora-duplicates & 4k & STS & \url{https://huggingface.co/datasets/sentence-transformers/quora-duplicates} \\

XNLI & 96k & NLI & \url{https://huggingface.co/datasets/mteb/xnli} \\

All-NLI & 145k & NLI & \url{https://huggingface.co/datasets/sentence-transformers/all-nli} \\

Signal1m-generated-queries & 200k\textsuperscript{*} & Twitter & \url{https://huggingface.co/datasets/BeIR/signal1m-generated-queries} \\

% General in total & 726,549 & -- & -- \\

\hline

\textbf{DM-MTT} & 412k \\

\hline
\textbf{Total} & 1.14M \\

\hline
\end{tabular*}
\caption{Composition of \textbf{MTT} used for supervised instruction fine-tuning in Stage~3. \textsuperscript{*}: After applying false positive filtering, we randomly sampled 200K examples from the remaining dataset. A portion of subsets in GD-MTP undergo hard negative mining (\S\ref{sec: hard_neg_mine}) to get GD-MTT.}

\label{tab:FT_data}
\end{table*}

\begin{table*}[!tp]
\centering
\small
\renewcommand{\arraystretch}{1.1}
\begin{tabular*}{\textwidth}{@{\extracolsep{\fill}}@{\hspace{5pt}}l l l l c c@{\hspace{5pt}}}
\hline

\textbf{Model}    & \textbf{Param}   & \textbf{Size}  & \textbf{Base}   & \textbf{Embed.} & \textbf{Arch.} \\

\textbf{Name}     & \textbf{Size}    & \textbf{Bin}   & \textbf{Model}  & \textbf{Size}   &               \\
\hline
Qwen3-Embedding-8B & 7.6B & XL & Qwen3-8B & 4096 & decoder\\
Qwen3-Embedding-4B & 4B & XL & Qwen3-4B & 2560 & decoder\\
inf-retriever-v1                       & 7B    & XL     & gte-Qwen2-7B-instruct               & 3584 & decoder \\
NV-Embed-v2                           & 7B    & XL     & Mistral-7B-v0.1                     & 4096 & decoder \\
Qwen3-Embedding-0.6B & 0.6B & Large & Qwen3-0.6B & 1024 & decoder \\
inf-retriever-v1-1.5b                  & 1.5B  & Large     & gte-Qwen2-1.5B-instruct             & 1536 & decoder \\
Linq-Embed-Mistral          & 7B    & XL     & E5-mistral-7b-instruct              & 4096 & decoder \\
NV-Embed-v1                           & 7B    & XL     & Mistral-7B-v0.1                     & 4096 & decoder \\
SFR-Embedding-Mistral             & 7B    & XL     & E5-mistral-7b-instruct              & 4096 & decoder \\
snowflake-arctic-embed-l           & 335M  & Medium & e5-large-unsupervised      & 1024 & encoder \\
snowflake-arctic-embed-l-v2.0      & 568M  & Large  & gte-multilingual-mlm-base           & 1024 & encoder \\
snowflake-arctic-embed-m-v2.0      & 305M  & Medium & bge-m3-retromae                     & 768  & encoder \\
gte-Qwen2-7B-instruct            & 7B    & XL     & Qwen2-7B                            & 3584 & decoder \\
snowflake-arctic-embed-m-v1.5      & 109M  & Small  &  BERT-base-uncased                                   & 768     & encoder \\
e5-mistral-7b-instruct              & 7B    & XL     & Mistral-7b                          & 4096 & decoder \\
snowflake-arctic-embed-m           & 109M  & Small  & e5-unsupervised-base       & 764  & encoder \\
snowflake-arctic-embed-m-long           & 137M  & Medium  & nomic-embed-text-v1-uns       & 768  & encoder \\

bge-large-en-v1.5                       & 335M  & Medium & --                                 & 1024 & encoder \\
mxbai-embed-large-v1           & 335M  & Medium &  --                                   & 1024 & encoder \\
snowflake-arctic-embed-s           & 33M   & Small  & e5-unsupervised-small      & 384  & encoder \\
bge-base-en-v1.5                        & 109M  & Small  & --                                  & 768  & encoder \\
bge-small-en-v1.5                       & 33M   & Small  & --                                  & 384  & encoder \\
multilingual-e5-large-instruct      & 560M  & Large  & xlm-roberta-large                   & 1024 & encoder \\
thenlper-gte-base                            & 109M  & Small  & EBRT-base                           & 768  & encoder \\
multilingual-e5-large                        & 560M  & Large  & xlm-roberta-large                                    & 1024 & encoder \\

thenlper-gte-small                            & 33M  & Small  & MiniLM-L12-H384                           & 384  & encoder \\

gte-Qwen2-1.5B-instruct          & 1.5B  & Large     & Qwen2-1.5B                          & 1536 & decoder \\

gte-base-en-v1.5                 & 137M  & Medium & EBRT-base                           & 768  & encoder \\
gte-large-en-v1.5                & 434M  & Large  & EBRT-large                          & 1024 & encoder \\
\hline
\end{tabular*}
\caption{Information of all baseline models. ``--'' means no publicly available information is available.}
\label{tab:model_specs}
\end{table*}

\begin{table*}[ht]
\centering
\small
\renewcommand{\arraystretch}{1.1}
\begin{tabular*}{\textwidth}{@{\extracolsep{\fill}}@{\hspace{5pt}}l p{0.45\textwidth} r@{\hspace{5pt}}}
\hline
\textbf{Model Name} & \textbf{Link} & \textbf{License} \\
\hline
Qwen3-Embedding-8B & \url{https://huggingface.co/Qwen/Qwen3-Embedding-8B} & apache-2.0 \\
Qwen3-Embedding-4B & \url{https://huggingface.co/Qwen/Qwen3-Embedding-4B} & apache-2.0 \\
inf-retriever-v1                       & \url{https://huggingface.co/infly/inf-retriever-v1} & apache-2.0 \\
NV-Embed-v2                            & \url{https://huggingface.co/nvidia/NV-Embed-v2} & cc-by-nc-4.0 \\
Qwen3-Embedding-4B & \url{https://huggingface.co/Qwen/Qwen3-Embedding-0.6B} & apache-2.0 \\
inf-retriever-v1-1.5b                  & \url{https://huggingface.co/infly/inf-retriever-v1-1.5b} & apache-2.0 \\
Linq-Embed-Mistral                     & \url{https://huggingface.co/Linq-AI-Research/Linq-Embed-Mistral} & cc-by-nc-4.0 \\
NV-Embed-v1                            & \url{https://huggingface.co/nvidia/NV-Embed-v1} & cc-by-nc-4.0 \\
SFR-Embedding-Mistral                  & \url{https://huggingface.co/Salesforce/SFR-Embedding-Mistral} & cc-by-nc-4.0 \\
snowflake-arctic-embed-l               & \url{https://huggingface.co/Snowflake/snowflake-arctic-embed-l} & apache-2.0 \\
snowflake-arctic-embed-l-v2.0          & \url{https://huggingface.co/Snowflake/snowflake-arctic-embed-l-v2.0} & apache-2.0 \\
snowflake-arctic-embed-m-v2.0          & \url{https://huggingface.co/Snowflake/snowflake-arctic-embed-m-v2.0} & apache-2.0 \\
gte-Qwen2-7B-instruct                  & \url{https://huggingface.co/Alibaba-NLP/gte-Qwen2-7B-instruct} & apache-2.0 \\
snowflake-arctic-embed-m-v1.5          & \url{https://huggingface.co/Snowflake/snowflake-arctic-embed-m-v1.5} & apache-2.0 \\
e5-mistral-7b-instruct                 & \url{https://huggingface.co/intfloat/e5-mistral-7b-instruct} & mit \\
snowflake-arctic-embed-m               & \url{https://huggingface.co/Snowflake/snowflake-arctic-embed-m} & apache-2.0 \\

snowflake-arctic-embed-m               & \url{https://huggingface.co/Snowflake/snowflake-arctic-embed-m-long} & apache-2.0 \\

granite-embedding-125m-english         & \url{https://huggingface.co/ibm-granite/granite-embedding-125m-english} & mit \\
bge-large-en-v1.5                      & \url{https://huggingface.co/BAAI/bge-large-en-v1.5} & apache-2.0 \\

snowflake-arctic-embed-s               & \url{https://huggingface.co/Snowflake/snowflake-arctic-embed-s} & mit \\

bge-base-en-v1.5                       & \url{https://huggingface.co/BAAI/bge-base-en-v1.5} & mit \\
bge-small-en-v1.5                      & \url{https://huggingface.co/BAAI/bge-small-en-v1.5} & mit \\
multilingual-e5-large-instruct         & \url{https://huggingface.co/intfloat/multilingual-e5-large-instruct} & mit \\
\
thenlper-gte-base                      & \url{https://huggingface.co/thenlper/gte-base} & mit \\

multilingual-e5-large                  & \url{https://huggingface.co/intfloat/multilingual-e5-large-instructt} & apache-2.0 \\
thenlper-gte-small                      & \url{https://huggingface.co/thenlper/gte-small} & mit \\

gte-Qwen2-1.5B-instruct                & \url{https://huggingface.co/Alibaba-NLP/gte-Qwen2-1.5B-instruct} & mit \\

gte-base-en-v1.5                       & \url{https://huggingface.co/Alibaba-NLP/gte-base-en-v1.5} & apache-2.0 \\
gte-large-en-v1.5                      & \url{https://huggingface.co/Alibaba-NLP/gte-large-en-v1.5} & mit \\
\hline
\end{tabular*}
\caption{HuggingFace model links and licenses for all evaluated models.}
\label{tab:model_links}
\end{table*}

\begin{figure*}[t]  % t=顶部, h=当前位置, b=底部, 你可以换
    \centering
    \includegraphics[width=\textwidth]{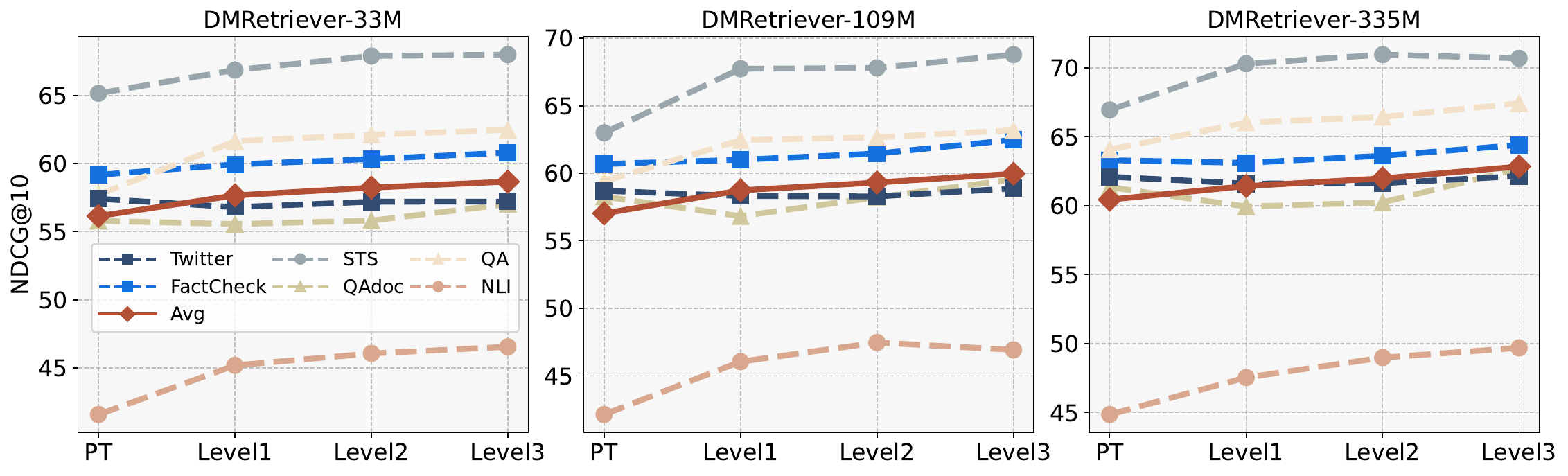}
    \caption{Model performance during different iterations of progressively supervised instruction fine-tuning}
    \label{fig:performance_stage}
\end{figure*}

\begin{figure*}[t] 
    \centering
    \includegraphics[width=\textwidth]{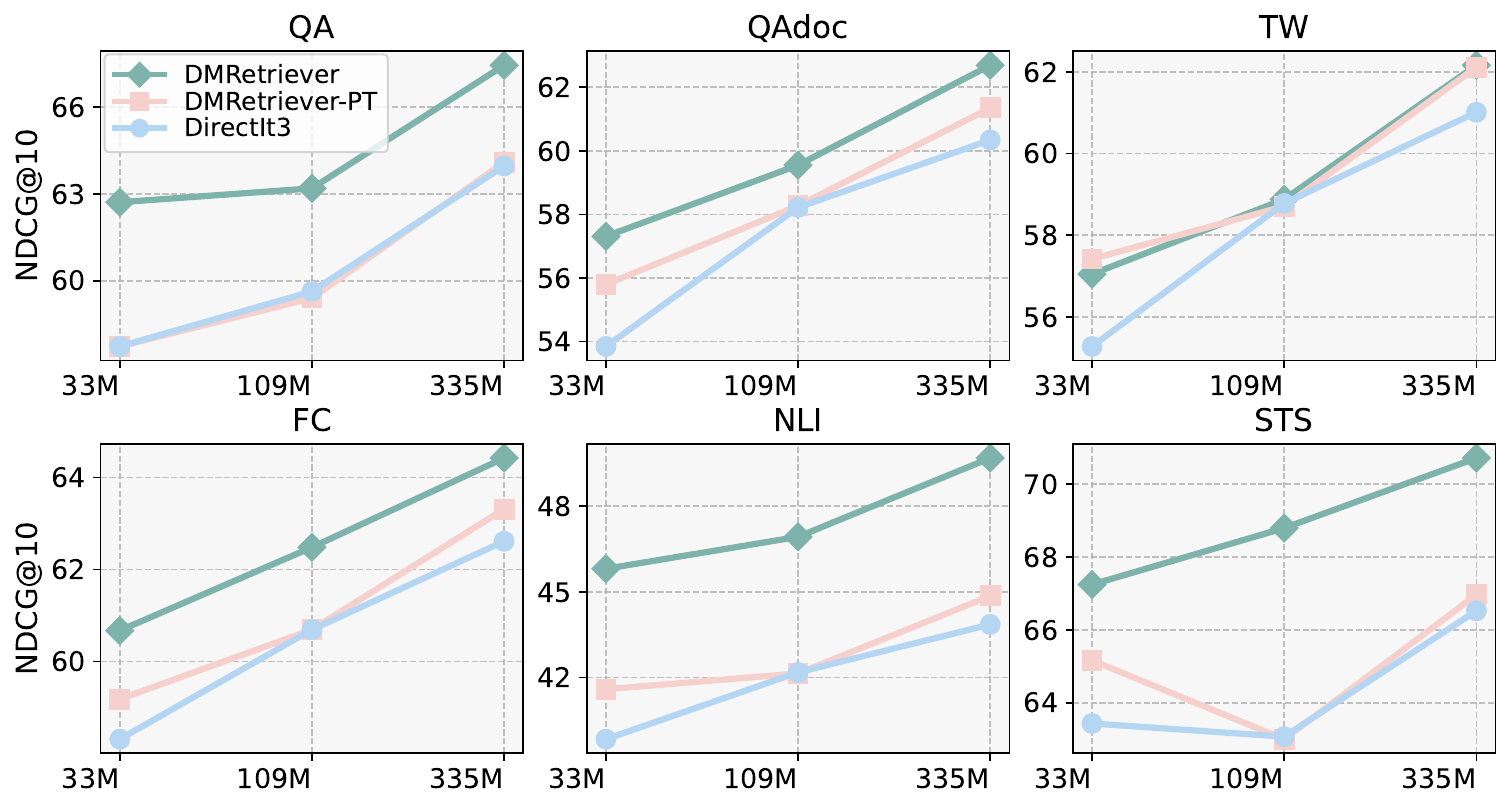}
    \caption{Comparisons of model performance under progressive fine-tuning and direct iteration-3 fine-tuning across 6 search intents.}
    \label{fig:Ablation_Stage_eachtask_FT}
\end{figure*}

\begin{figure*}[t]
    \centering
    \includegraphics[width=\textwidth]{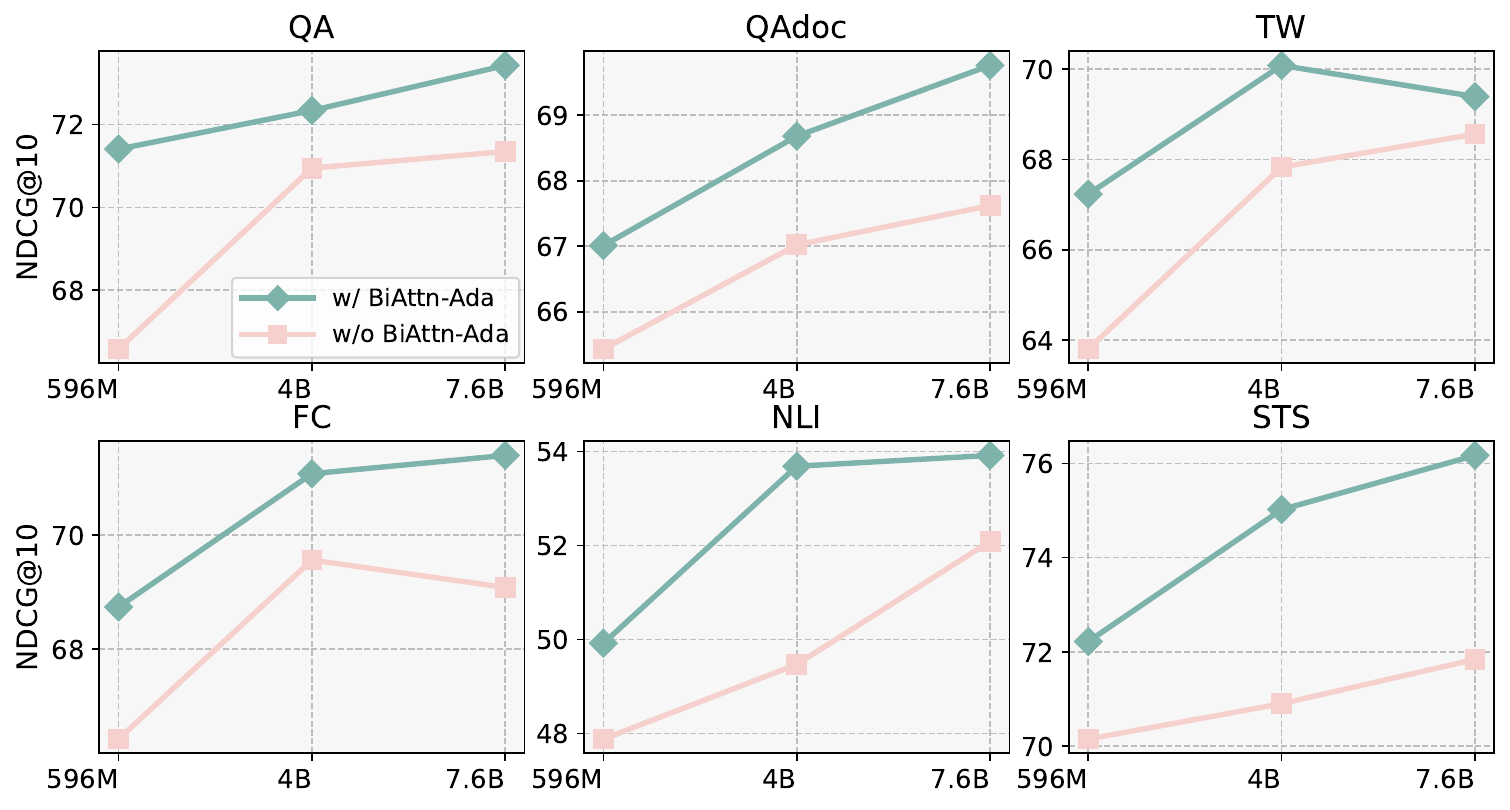}
    \caption{Model performance comparisons between with and without bidirectional attention adaptation across 6 search intents.}
    \label{fig:Ablation_Stage_eachtask_PTS1}
\end{figure*}

\begin{figure*}[t] 
    \centering
    \includegraphics[width=\textwidth]{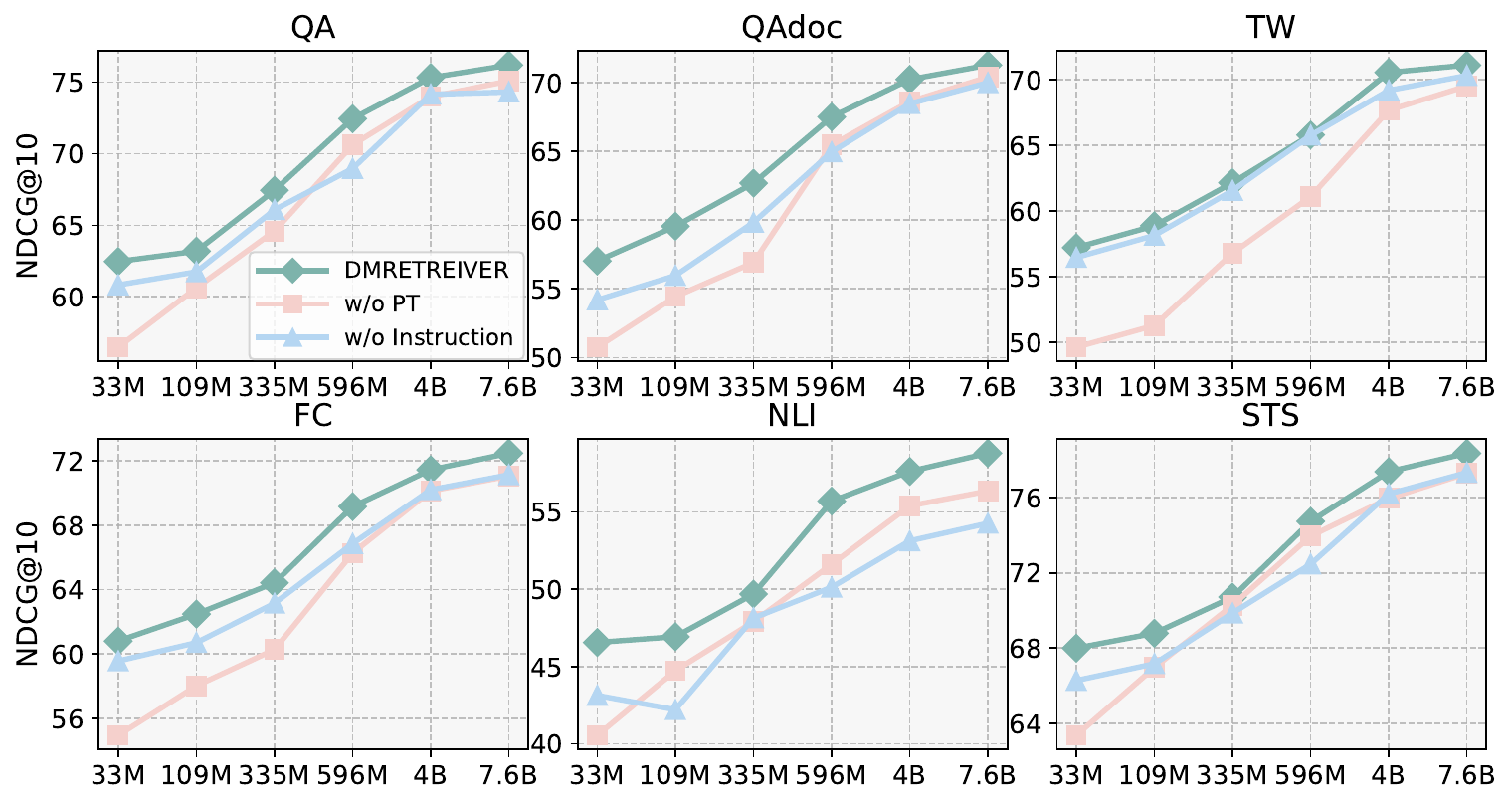}
    \caption{Model performance under ablations of whether conducting pre-training and adding instruction across 6 search intents.}
    \label{fig:Ablation_Stage_eachtask_PT_vs_noPT}
\end{figure*}

\end{document}